\documentclass[preprint,aps,prd,showpacs,nofootinbib,superscriptaddress,floatfix]{revtex4-1}
\usepackage{amsmath}
\usepackage{graphicx}
\usepackage{subfigure}
\usepackage{epstopdf}
\usepackage{epsfig}
\usepackage{amssymb}
\usepackage{bm}
\usepackage{bbm}
\newcommand{\beq}{\begin{equation}}
\newcommand{\eeq}{\end{equation}}

\usepackage{color}

\begin{document}

\title{ 
Chiral $\bm{U(1)}$ flavor models and flavored Higgs doublets:
\\
the top FB asymmetry and the $\bm{Wjj}$
}

\author{P. Ko}
\email[]{pko@kias.re.kr}
\affiliation{School of Physics, KIAS, Seoul 130-722, Korea}

\author{Yuji Omura}
\email[]{omura@kias.re.kr}
\affiliation{School of Physics, KIAS, Seoul 130-722, Korea}

\author{Chaehyun Yu}
\email[]{chyu@kias.re.kr}
\affiliation{School of Physics, KIAS, Seoul 130-722, Korea}


\begin{abstract}
\noindent
We present $U(1)$ flavor models for leptophobic $Z^{'}$ with flavor dependent
couplings to the right-handed up-type quarks in the Standard Model,
which can accommodate the recent data on the top forward-backward (FB)
asymmetry and the dijet resonance associated with a $W$ boson
reported by CDF Collaboration.
Such flavor-dependent leptophobic charge assignments generally require
extra chiral fermions for anomaly cancellation. 
Also the chiral nature of $U(1)'$ flavor symmetry calls for 
new $U(1)'$-charged Higgs doublets in order for the SM fermions to have 
realistic renormalizable Yukawa couplings.
The stringent constraints from the top FB asymmetry at the Tevatron
and the same sign top pair production at the LHC can be evaded
due to contributions of the extra Higgs doublets.
We also show that the extension could realize cold dark matter candidates.
\end{abstract}


\maketitle

\section{Introduction}
\label{sec:intro}

The top forward-backward (FB) asymmetry ($A_{\rm FB}^t$) 
measured at the Tevatron has drawn a lot of attention during the past few years.  
The most recent updated data from CDF and D0 are  
\cite{CDFAFB1,CDFAFB2,D0AFB}
\begin{equation}
A_{\rm FB}^t = \left\{ \begin{array}{cl}
0.158 \pm 0.074 & (\textrm{CDF, lepton+jets channel}) \\
0.42 \pm 0.158 & (\textrm{CDF, dilepton channel}) \\
0.19  \pm 0.065 & (\textrm{D0, lepton+jets channel})
\end{array}   \right.
\label{eq:afb_cdf}
\end{equation}
in the $t\bar{t}$ rest frame, whereas 
the SM prediction \cite{Kuhn:1998jr,Kuhn:1998kw,Bowen:2005ap,Antunano:2007da}
based on MCFM is $0.058 \pm 0.009$ \cite{CDFAFB1}. 
Recent calculations of the top FB asymmetry at the next-next-leading-log 
(NNLL)  do not differ much from the aforementioned SM predictions, 
and the sizable discrepancy still remains \cite{Ahrens:2011mw}.

Motivated by this discrepancy, numerous suggestions have been 
made on how to explain the observed large top FB asymmetry 
\cite{Choudhury:2007ux,Jung:2009jz,Cheung:2009ch,Shu:2009xf,Arhrib:2009hu,%
Dorsner:2009mq,Jung:2009pi,Jung:2011ym,%
Barger:2010mw,Cao:2010zb,Xiao:2010hm,Jung:2010yn,%
Choudhury:2010cd,Cheung:2011qa,Gresham:2011dg,%
Bhattacherjee:2011nr,Barger:2011ih,%
Grinstein:2011yv,Isidori:2011dp,Zerwekh:2011wf,Barreto:2011au,%
Foot:2011xu,Ligeti:2011vt,AguilarSaavedra:2011vw,Gresham:2011pa,Shu:2011au,%
AguilarSaavedra:2011zy,Krohn:2011tw,Cui:2011xy,Gabrielli:2011jf,%
Duraisamy:2011pt,Ahrens:2011uf,AguilarSaavedra:2011ug,Tavares:2011zg,%
Shao:2011wa,Blum:2011fa,Gresham:2011fx,Frank:2011rb,Davoudiasl:2011tv,%
Jung:2011id,Nelson:2011us,Jung:2011ua,Zhu:2011ww,Jung:2011ue,Babu,%
Hektor:2011ms,Vecchi}.
The models can be categorized into the following: 
colored spin-1 (axigluon, coloron, Kaluza-Klein gluon, etc.) exchange 
in the $s$-channel, light $Z^{'}$ or $W^{' \pm}$ 
exchange in the $t$-channel,  color antitriplet or sextet in the $u$-channel, 
color-singlet scalar exchange in the $t$-channel, 
and effective lagrangian approaches. 
Some models could be tested at the LHC. For example, the original 
light $Z^{'}$ model \cite{Jung:2009jz} is excluded
by the recent CMS data on the same sign top pair production \cite{CMSsametop}. 

However, most models are phenomenologically motivated by the top FB asymmetry.  
And the issues related with flavor dependent gauge symmetry,  
anomaly cancellation 
and renormalizable Yukawa couplings were not properly addressed. 
When one considers a complete model including new particles and interactions 
needed for the top FB asymmetry, there could be additional degrees of freedom
that might contribute to the top FB asymmetry. Therefore it may be premature to
conclude which model is favored or not.  This could be especially the case  
for models based on a new spin-1 particle.

Independent of the top FB asymmetry, the CDF Collaboration reported an 
interesting excess on the dijet production associated with a $W$ boson 
in a dijet mass range between 130 GeV and 160 GeV 
with an integrated luminosity of $4.3$ fb$^{-1}$ \cite{CDFWjj}.
The excess could be interpreted as the production process $p\bar{p}\to W X$
followed by $X \to jj$, where $\sigma(WX)\sim 4$ pb with $m_X\sim 145$ GeV.
A possible candidate for $X$ is a light $Z^\prime$ boson
\cite{Buckley:2011vc,Cheung:2011zt,Yu:2011cw,Ko:2011ns,%
Fox:2011qd,Jung:2011ua,Jung:2011ue}.
Various resolutions without invoking the $Z^\prime$ boson 
also have been  proposed to reconcile the CDF $Wjj$ excess
\cite{Nelson:2011us,Jung:2011ua,Zhu:2011ww,Jung:2011ue,Babu,%
Hektor:2011ms,Vecchi,%
Eichten:2011sh,He:2011ss,wang:2011taa,Sato:2011ui,Anchordoqui:2011ag,%
Sullivan:2011hu,Plehn:2011nx,Cao:2011yt,Chang:2011wj,Dutta:2011kg,%
Carpenter:2011yj,Chen:2011wp,Liu:2011di,Hewett:2011nb,Fan:2011vw,%
Gunion:2011bx,Cheung:2011vx,Ghosh:2011np,Wang:2011uq}.
The excess was confirmed by the CDF Collaboration with larger data, but
was not confirmed by the D0 Collaboration \cite{D0Wjj}.
It would be remained to be seen if the CDF $Wjj$ excess survives in the future,
but it would be desirable to calculate the typical size of $\sigma (Wjj)$  in 
new physics model under consideration.

Recently, the present authors proposed an extension of the Standard Model (SM) 
where flavor-dependent $U(1)'$ charges were assigned to the right-handed (RH)
up-type quarks \cite{Ko-Top}, and they made the light $Z^{'}$ model 
of Ref.~\cite{Jung:2009jz} for the top FB asymmetry complete and realistic 
by constructing full renormalizable and anomaly free models with flavor dependent 
charge assignments to the right-handed up-type quarks. 
In particular,  it was shown that the light $Z^{'}$ solution,  
which is now disfavored by the same sign top pair production constraint 
by the CMS Collaboration \cite{CMSsametop},  
could be revived because there are additional $t$-channel 
contributions from the neutral (pseudo) scalar Higgs bosons 
to the top FB asymmetry and the same sign top pair production. 
There is destructive interference among $Z^{'}$ 
and neutral (pseudo) scalar Higgs 
bosons in the latter observable,  making the light $Z^{'}$ scenario with light 
neutral (pseudo)scalar bosons still a viable solution to the top FB asymmetry.  
Such models suffer from the constraints from flavor changing neutral currents 
(FCNC) and the process involving top quarks, so that the charge assignments 
have to be controlled.
In Ref.~\cite{Ko-Top}, the authors concentrated on the cases that the nonzero 
charges are assigned only to the right-handed up-type quarks, and 
demonstrated that the models provide a possible resolution of the top 
forward-backward asymmetry \cite{CDFAFB1}, the CDF $Wjj$ 
excess~\cite{CDFWjj}, and cold dark matters (CDMs).
In this paper, we elaborate on these models considering various $U(1)^{'}$
charge assignments to the SM quarks, and including $U(1)^{'}$
charged Higgs doublets that are to be introduced in order that we can write
renormalizable Yukawa couplings for the SM fermions.

Flavor-dependent $U(1)'$ charge assignments would require the extensions 
of Higgs doublets in order to realize realistic mass matrices with 
renormalizable Yukawa couplings, if the SM fermions have chiral $U(1)'$ charges.
Since the right-handed up-type quarks  in the models of Ref.~\cite{Ko-Top} and 
this paper have flavor dependent couplings whereas other fermions have 
flavor universal couplings, the SM fermions 
(at least the right-handed up-type quarks) 
have chiral $U(1)'$ charges so that we have to introduce $U(1)'$ charged Higgs 
doublets 
in order to write the realistic Yukawa couplings for all the SM fermions
and generate their masses. As pointed out in Ref. \cite{Blum:2011fa,Babu}, 
such extra $SU(2)_L$ Higgs doublets may enhance the $A_\textrm{FB}$.  
In our model, it turns out that the top FB asymmetry and the same sign top pair
production receive contributions not only from the $U(1)'$ gauge boson, $Z'$, 
but also from neutral scalar Higgs, pseudo-scalar Higgs, and charged Higgs. 
As we see in the Sec.\ref{sec:phenomenology}, such Higgs contributions play 
an important role in achieving the favored region for the top FB asymmetry 
without conflict with the same sign top pair production, and can accommodate
the CDF $Wjj$ signal when the excess is confirmed in the future.

Our models would not be anomaly-free without extra chiral fermions.
One simple way to realize anomaly-free theory is adding one extra generation 
and two SM gauge vector-like fermion pairs. The added fields may also 
contribute to observed signals and raise new predictions.
In fact, FCNC problems will be triggered in some cases, but stable particles, 
which become CDM candidates, will be guaranteed 
because of $U(1)'$ symmetry.    

On the other hand, such flavor-dependent $U(1)'$ symmetric models may be known
as the Froggatt-Nielsen Model (FN) \cite{FN}, where the Yukawa terms are 
expressed by the power of a SM gauge singlet, $\Phi$. 
This model is not renormalizable, but $Z'$ and neutral scalar Higgs fields 
can contribute to $A_{FB}$.
We also comment on the possibility that it may be 
compatible with the large $A_\textrm{FB}$ to realize the hierarchical 
structures of Yukawa textures according to the power of $\Phi$ like FN, 
in Sec. \ref{sec:FN-model}.

This paper is organized as follows. 
In Sec. \ref{sec:model} A and B,  we describe the flavor dependent 
leptophobic $U(1)^{'}$ models, gauge and Yukawa interactions, 
respectively. In Sec. \ref{sec:anomaly}, 
we discuss the conditions for anomaly cancellation and 
introduce extra chiral fields for the anomaly cancellation.
In Sec. \ref{sec:FN-model}, 
we discuss the FN-type model where Yukawa couplings are expressed 
by higher-order terms, and then we discuss the explicit models 
with extra $SU(2)_L$ Higgs doublets charged under $U(1)'$, 
in Sec.~\ref{sec:multi-Higgs-exam}, based on the argument 
in Sec.~\ref{sec:multi-Higgs}: one example is two Higgs doublet model
and the other is three Higgs doublet model. 
In Sec. \ref{sec:phenomenology}, we discuss phenomenology of 
each model, and describe the contribution of $Z^\prime$ and Higgs bosons 
to $A_\textrm{FB}$ and the same sign top pair production at the Tevatron
and LHC. We also show that the $Wjj$ excess reported 
by CDF might be interpreted as a leptophobic $Z'$ through
$p\bar{p}\to h^\pm \to W^\pm Z^\prime$ followed by $Z^\prime \to jj$.
In Sec. \ref{sec:CDM}, we comment on how to achieve stable particles, 
which are good CDM candidates, in our models, and Sec. \ref{sec:summary} 
is devoted to summary.  In the appendix, we show the explicit descriptions of 
Yukawa couplings with Higgs and quarks in the three-Higgs models.  

\section{$\bm{U(1)'}$ Flavor Models with $\bm{U(1)'}$-charged Higgs doublets}
\label{sec:model}

\subsection{$\bm{U(1)^{'}}$ Gauge Interactions}

In this work, we are interested in explaining the top FB asymmetry 
in terms of relatively light $Z^{'}$ with mass around 150 GeV. 
Such a light $Z'$ should be leptophobic in order to evade the stringent
bound from Drell-Yan processes from the Tevatron and the LHC.  
Therefore we start with leptophobic $U(1)'$ gauge models with the 
following flavor-dependent charge assignments: 
\begin{center}
\begin{tabular}{|c|c|c|c|c|c|c|}\hline
          & $SU(3)_c$ & $SU(2)_L$ & $U(1)_Y$ & $U(1)'$    \\ \hline  
$Q_i$     & $3$       & $2$       & $1/6$    & $q_{i}$    \\ \hline 
$D_{Ri}$  & $3$       & $1$       & $-1/3$   & $d_i$      \\ \hline
$U_{Ri}$  & $3$       & $1$       & $2/3$    & $u_i$      \\ \hline  
$L_i$     & $1$       & $2$       & $-1/2$   & $0$        \\ \hline 
$E_{Ri}$  & $1$       & $1$       & $-1$     & $0$        \\ \hline
$H$       & $1$       & $2$       & $1/2$    & $q_h$      \\ \hline 
$\Phi$    & $1$       & $1$       & $0$      & $q_{\Phi}$ \\ \hline  
\end{tabular}
\end{center}
with $Q_i^T =(U_{Li},\,\! D_{Li})$, 
$L_i^T =(\nu_{Li},\,\! E_{Li})$  and $i=1,2,3$ are generation indices.  
$\Phi$ is a SM-gauge singlet with nonzero $U(1)'$ charge which is 
required to break $U(1)'$ spontaneously and 
generate nonzero mass for $Z'$.  

Flavor-dependent extra $U(1)'$ models were very popular 
in order to explain the origin of Yukawa texture in the SM in terms of 
flavor dependent $U(1)'$ charges of the SM fermions.  
For example, Froggatt and Nielsen define each fermion charges 
corresponding to the mass hierarchy, and explain the small Yukawa 
couplings using the suppression from higher-order terms~\cite{FN}.
This type of models for flavor has serious conflict with the low energy
constraints on the highly suppressed FCNC, if the $U(1)'$ gauge boson
$Z'$ is as low as $\sim 150$ GeV.  After all, it would be very difficult to 
accommodate the observed top FB asymmetry in the FN framework, 
the detailed arguments for which will be given in the later subsection
(see Sec.~\ref{sec:FN-model}).

Let us define the couplings between $Z'$ and the SM quarks
in the interaction eigenstates and in the mass (flavor) eigenstates: 
\begin{eqnarray}
{\cal L}_{Z' f \bar{f}} & = & g' Z'_{\mu}   
\left[  ~ q_i \overline{U^i_L} \gamma^{\mu} 
U^i_L + q_i \overline{D^i_L} \gamma^{\mu} D^i_L 
+ u_i \overline{U^i_R}  \gamma^{\mu} U^i_R + 
d_i \overline{D^i_R} \gamma^{\mu} D^i_R \right]
\nonumber
\\ 
& = & g' Z'_{\mu} 
\left[  (g^u_L)_{ij} \overline{\Hat{U}^i_L} \gamma^{\mu} 
\Hat{U}^j_L +(g^d_L)_{ij} \overline{\Hat{D}^i_L} \gamma^{\mu} \Hat{D}^j_L 
+ (g^u_R)_{ij} \overline{\Hat{U}^i_R} \gamma^{\mu} 
\Hat{U}_R^j +(g^d_R)_{ij} \overline{\Hat{D}^i_R} \gamma^{\mu} 
\Hat{D}^j_R \right] .
\end{eqnarray}
Sum over the repeated indices are implicitly understood.
The Yukawa matrices and the mass matrices of up- and down-type 
quarks are related as
\[ 
Y^u_{ij} = (L_u)^{\dagger}_{ik} m_k^{u}  (R_u)_{kj} , \ \ \ 
Y^d_{ij} = (L_d)^{\dagger}_{ik} m_k^{d}  (R_d)_{kj} .
\]
Then the $U(1)'$ couplings of up- and down-type quarks are given by 
$g^u_{L,R}$ and $g^d_{LR}$:  
\begin{eqnarray}
\label{eq:gLgR}
(g^u_L)_{ij} &=&  (L_u)_{ik}  q_{k} (L_u)^{\dagger}_{kj},~ 
(g^d_L)_{ij} = (L_d)_{ik}  q_{k} (L_d)^{\dagger}_{kj}, 
\nonumber \\  
(g^u_R)_{ij} &=&(R_u)_{ik}  u_k (R_u)^{\dagger}_{kj},~ 
(g^d_R)_{ij} =  (R_d)_{ik}  d_k (R_d)^{\dagger}_{kj}.
\end{eqnarray}

One remark is in order. When one introduces a flavor dependent $U(1)'$
for phenomenologically motivated reasons, it is important to define 
the $U(1)'$ charges of the SM fermions before electroweak symmetry 
breaking (EWSB) because the fermion flavor is defined 
in terms of mass eigenstates 
only after the EWSB.  It would be not reasonable to assume that physical 
$Z'$ has a nonzero flavor changing interactions in a particular channel 
in the fermion mass eigenstates. It would be more natural to assume a
particular $U(1)'$ assignments in interaction eigenstates. Then this particular
charge assignments will be (partly) erased when we go to the mass eigenstates
using Eq.~(\ref{eq:gLgR}).

In the left-handed quark sector, a sizable mixing is required to realize 
CKM matrix. Therefore the flavor changing interactions 
in the down quark sector,  
especially $(1,2)$-element, would be problematic if $Z'$ mass is light.   
As far as we know, the data on 
$K^0-\overline{K^0}$, $B_d^0-\overline{B_d^0}$, 
$B_s^0 - \overline{B_s^0}$, and $D^0 - \overline{D^0}$
can be accommodated 
within the SM CKM paradigm very well,  so that it might be difficult to assign 
flavor-dependent charges to, especially, down-type quarks (both $D_{Li}$ 
and $D_{Ri}$).  This phenomenological constraints can be easily accommodated
if we assume that the left-handed up- and down-type quarks and the RH 
down-type quarks have flavor universal $U(1)'$ charges 
(including null charge). 
Then the usual GIM mechanism becomes operative in the down quark sector, 
and one can easily verify that the flavor changing couplings are allowed only 
in the RH up-type sector.  This would be exactly what we need for explaining 
the top FB asymmetry in terms of light $Z'$.  
The $D^0 - \overline{D^0}$ mixing would be controlled as described in 
the Sec. \ref{sec:multi-Higgs-exam},  and we will focus mainly on $t-u$ mixing 
which may evade bounds from collider physics.
\footnote{For example, see Ref.~\cite{Blum:2011fa, Gedalia:2009kh, 
Blum:2009sk} about the FCNC bounds.} 
In the subsection \ref{sec:multi-Higgs}, we introduce the models 
with extra $U(1)'$-charged Higgs doublets which realize this scenario
evading  the stringent constraints from the highly suppressed FCNC.  

\subsection{Renormalizable Yukawa Interactions calls 
for $\bm{U(1)'}$-charged Higgs doublets}
\label{sec:multi-Higgs}
Here, we introduce renormalizable models which can evade the strong 
FCNC bounds. We assume that the left-handed up- and down-type quarks and 
right-handed down-type quarks have universal charge, and only charges of 
right-handed up-type quarks are flavor-dependent:   
\begin{center}
\begin{tabular}{|c|c|c|c|c|c|c|}\hline
          & $SU(3)_c$ & $SU(2)_L$ & $U(1)_Y$ & $U(1)'$      \\ \hline  
$Q_i$     & $3$       & $2$       & $1/6$    & $q_L$        \\ \hline 
$D_{Ri}$  & $3$       & $1$       & $-1/3$   & $q_L$        \\ \hline
$U_{Ri}$  & $3$       & $1$       & $2/3$    & $u_i$        \\ \hline  
\end{tabular}
\end{center}
The above charge assignment realizes
\begin{equation}
 (g^u_L)_{ij} =(g^d_L)_{ij}=(g^d_R)_{ij} =q_L\delta_{ij}, 
\end{equation}
which can avoid dangerous tree-level FCNC 
contributions except for the right-handed up quark sector.  
The mixing matrix $(g^u_R)_{ij}$ have nonzero off-diagonal elements, 
so that we could discuss the cases that $(g^u_R)_{ut}$ is large and 
$(g^u_R)_{uc}$, which contributes to $D^0 - \overline{D^0}$, 
is suppressed, adopting appropriate Yukawa couplings. 

In order to achieve realistic mass matrices and renormalizability 
in the models with flavor-dependent charge assignments,  
more than one extra Higgs doublet, charged under $U(1)'$, are required.  
For example, the assignment satisfying 
$u_1=u_2=q_L$ and $u_2 \neq u_3$ requires at least one extra Higgs 
doublet field whose  $U(1)'$ charge is $(-q_L +u_3)$. 
In the most generic case $u_1 \neq u_2 \neq u_3 \neq u_1$, 
three extra Higgs doublets are required for renormalizable Yukawa couplings.  
If one of $u_i$ is define as $u_i= q_L$, 
two extra Higgs doublet fields are necessary for realistic Yukawa matrix.   
In such multi-Higgs models, we 
have not only neutral scalar Higgs fields, but also pseudoscalar and charged 
Higgs fields.  If their masses are around weak scale,
they may also contribute to flavor changing processes through Yukawa couplings.
In Sec.~\ref{sec:multi-Higgs-exam}, we investigate $2$ explicit models, 
setting $q_L$ to $0$: 
one is $(u_1,\,\! u_2,\,\! u_3)=(0,\,\! 0,\,\! 1)$ assignment
which corresponds to $2$ Higgs doublet model (2HDM), 
and the other is $(u_1,\,\! u_2,\,\! u_3)=(-1,\,\! 0,\,\! 1)$ assignment
which corresponds to $3$ Higgs doublet model (3HDM).

\subsection{Conditions for anomaly cancellation}
\label{sec:anomaly}

Leptophobic and flavor-dependent $U(1)'$ models generally become 
anomalous without extra chiral fields.
One of the simplest ways to construct anomaly-free theory is adding 
one extra generation and two SM gauge vector-like pairs as follows:
\begin{center}
\begin{tabular}{|c|c|c|c|c|c|c|}\hline
         & $SU(3)_c$ & $SU(2)_L$ & $U(1)_Y$ & $U(1)'$                   \\ 
\hline  
$Q'$     & $3$       & $2$       & $1/6$    & $-(q_{1}+q_{2}+q_{3})$ \\
\hline 
$D'_R$   & $3$       & $1$       & $-1/3$   & $-(d_1+d_2+d_3)$          \\ 
\hline 
$U'_R$   & $3$       & $1$       & $2/3$    & $-(u_1+u_2+u_3)$\\ \hline 
$L'$     & $1$       & $2$       & $-1/2$   & $0$             \\ \hline 
$E'$     & $1$       & $1$       & $-1$     & $0$             \\ \hline  
$l_{L1}$ & $1$       & $2$       & $-1/2$   & $Q_L$           \\ \hline 
$l_{R1}$ & $1$       & $2$       & $-1/2$   & $Q_R$           \\ \hline 
$l_{L2}$ & $1$       & $2$       & $-1/2$   & $-Q_L$          \\ \hline 
$l_{R2}$ & $1$       & $2$       & $-1/2$   & $-Q_R$          \\ \hline 
\end{tabular}
\end{center}
One can replace the $SU(2)_L$ doublets $(l_{Li},\,\! l_{Ri})$ with $SU(3)_c$ 
triplets.\footnote{One can also replace them with fields charged 
or not charged under $SU(3)_c \times SU(2)_L$. }
\begin{center}
\begin{tabular}{|c|c|c|c|c|c|c|}\hline
         & $SU(3)_c$ & $SU(2)_L$ & $U(1)_Y$ & $U(1)'$    \\ \hline  
$q_{L1}$ & $3$       & $1$       & $-1/3$   & $Q_L$      \\ \hline 
$q_{R1}$ & $3$       & $1$       & $-1/3$   & $Q_R$      \\ \hline 
$q_{L2}$ & $3$       & $1$       & $-1/3$   & $-Q_L$     \\ \hline 
$q_{R2}$ & $3$       & $1$       & $-1/3$   & $-Q_R$     \\ \hline  
\end{tabular}
\end{center}
$Q_{L,R}$, $q_{i}$, $d_i$ and $u_i$ must satisfy the following two equations 
for the $U(1)'^3$ and $U(1)_YU(1)'^2$ anomaly cancellation, 
\begin{equation}
\label{eq:anomaly1}
6 \sum_i q_{i}^3 -6 \left( \sum_i q_{i} \right)^3-3 \sum_i d_{i}^3 
+3 \left( \sum_i d_{i} \right)^3-3 \sum_i u_{i}^3 
+3 \left( \sum_i u_{i} \right)^3=0,
\end{equation}
\begin{equation}
\label{eq:anomaly2}
\sum_i q_{i}^2 + \left( \sum_i q_{i} \right)^2+ \sum_i d_{i}^2 
+ \left( \sum_i d_{i} \right)^2-2 \sum_i u_{i}^2 
-2 \left( \sum_i u_{i} \right)^2=2(Q_L^2-Q_R^2).
\end{equation}
If one requires rational numbers for all the $U(1)'$ charges, 
there might be not so many simple solutions of Eq. (\ref{eq:anomaly1}). 
In the Sec. \ref{sec:multi-Higgs-exam}, we discuss concrete models
 satisfying Eq. (\ref{eq:anomaly1}) and Eq. (\ref{eq:anomaly2}).

We also have constraints on the $U(1)'$-charges from the Yukawa couplings 
which generate masses of the SM fermions and extra fermions.
Furthermore, our charge assignment may allow the mixing between the SM fermions 
and the extra fermions, which causes the FCNC problems. 
For example, in the case that $q_h$ is set to $0$,
we have the mixing terms between SM leptons and extra leptons,
\begin{equation}
\lambda^l_j
\overline{L' } H E_j +\lambda^E_i  \overline{L_i} H  E'  +h.c.
\end{equation}
We have to assume that $\lambda^{l,E}_j$ are controlled to avoid the problem. 
In fact, in the Sec.~\ref{sec:3-Higgs}, we consider the case 
with $q_{i}=d_i=0$ and $\sum_i u_{i}=0$ where the extra generation is 
not required and the SM gauge vector-like pairs do not have the mixing.

\subsection{Comments on the FN models}
\label{sec:FN-model}%

Before the introduction of our explicit models, let us comment on FN-type models.
The FN framework~\cite{FN} based on some $U(1)$ or other 
flavor symmetries  broken at some high energy scale has been 
a very popular way to understand the flavor structures of Yukawa matrices.  
One can easily admit that this is probably best motivation for considering 
flavor dependent gauge symmetry, whether it may be abelian or nonabelian. 
Froggatt and Nielsen define each fermion charges 
in order to explain the mass hierarchy and flavor mixings.  
The small Yukawa couplings are achieved by the suppression from 
higher-order terms~\cite{FN}.

More explicitly, let us introduce a FN model without extra 
$SU(2)_L$ Higgs doublets charged under $U(1)'$. 
Now, the left-handed and RH down-type quarks have 
flavor-dependent charges for the Yukawa texture.
Then Yukawa couplings are given by higher dimensional operators 
involving some powers of $\Phi$, which is SM gauge singlet with 
$U(1)'$ charge, $q_{\Phi}$,
\begin{equation} 
\label{eq:Yukawa-for-one}%
{y_{ij}}^u \left( \frac{\Phi}{M} \right)^{n^u _{ij}}  \overline{Q_i} 
\widetilde{H} U_{Rj}  +{y_{ij}}^d \left( \frac{\Phi}{M} \right)^{n^d _{ij}}  
\overline{Q_i} H D_{Rj} +h.c.,
\end{equation} 
where $\widetilde{H} \equiv i \tau_2 H^*$ and $(n^u_{ij},\,\! n^d_{ij})=
((q_{i}-u_j+q_h)/q_{\Phi},\,\! (q_{i}-d_j-q_h)/q_{\Phi})$. $M$ is 
the cut-off scale, and ${y_{ij}}^{u,d}$ are coupling constants. 
After $\Phi$ gets the nonzero vacuum expectation value (vev),
effective Yukawa couplings, $Y_{ij} ^{u,d}$, are induced,
\begin{equation}
Y_{ij} ^u = y_{ij} ^u \epsilon^{q_{i}-u_j +q_h},~Y_{ij} ^d 
= y_{ij} ^d \epsilon^{q_{i}-d_j -q_h},  
\end{equation} 
where $\epsilon$ is defined as $\epsilon \equiv \langle \Phi \rangle / M$ 
and $q_{\Phi}$ is set to $1$.
If $\epsilon$ is around Cabibbo angle, $\sim 0.22$, we can expect 
that the Yukawa texture, which can realize mass hierarchy and small mixing, 
is appearing effectively, and the degree of the suppression may be described 
by the power of $\epsilon$, when $y^{u,d}_{ij}$ are set to around $1$. 
For example, assuming that $Y_{ij}^{u,d}$ are close to diagonal, 
we find that the ratio of masses and mixing are roughly estimated as  
\begin{equation}
\label{eq;mass-FN}
\frac{m^u_i}{m^u_j} \sim \epsilon^{q_{i}-q_{j}-u_i+u_j},
~\frac{m^d_i}{m^d_j} \sim \epsilon^{q_{i}-q_{j}-d_i+d_j},~(i<j),
\end{equation}
and
\begin{equation}
 (V_{\textrm{CKM}})_{ij}  \sim  \epsilon^{|q_{i}-q_{j}|}.
\end{equation}
The size of each $(L_{u,d})_{ij}$ and $(R_{u,d})_{ij}$ can be also 
majored by the power of $\epsilon$,
and especially $L_u$ and $L_d$ contribute to CKM matrix, $V_{\textrm{CKM}}$, 
so that each couplings are estimated as follows: $(g^{u,d}_L)_{ij}$ are
\begin{eqnarray}
\label{eq:gL}%
(g^u_L)_{ij} & \sim& q_{i} \delta_{ij} + O((V_{\textrm{CKM}})_{ij}) 
(q_{j}-q_{i}), 
\\ \nonumber
(g^d_L)_{ij} & \sim & q_{i} \delta_{ij} + O((V_{\textrm{CKM}})_{ij}) 
(q_{j}-q_{i}), 
\end{eqnarray}
where $(L_{u,d})_{ij}$ are estimated as $(V_{\textrm{CKM}})_{ij}$, and 
$(g^{u,d}_R)_{ij}$ are
\begin{eqnarray}
\label{eq:gR}%
(g^u_R)_{ij} & \sim& u_i \delta_{ij} + (\delta R_u)_{ij} (u_j-u_i), \\ \nonumber
(g^d_R)_{ij}  &\sim& d_i \delta_{ij} + (\delta R_d)_{ij} (d_j-d_i). 
\end{eqnarray}
$(\delta R_{u,d})_{ij}$ are the small mixing defined as 
$(R_{u,d})_{ij} = \delta_{ij}+(\delta R_{u,d})_{ij}$ and 
they are also estimated according to the mass ratio and CKM matrix,
\begin{eqnarray}
\label{eq;mix-FN}
|( R_u)_{ij}| &\sim&|( R_u)_{ji}| \sim \epsilon^{|u_i-u_j|}, \\ \nonumber
|(R_d)_{ij}| &\sim&|(R_d)_{ji}| \sim \epsilon^{|d_i-d_j|}. 
\end{eqnarray}
We notice that large flavor changing couplings are induced 
by large CKM elements and small mass hierarchy. 
For instance, $(L_d)_{12}$, which contributes to $K_0$-$\overline{K}_0$ 
mixing, is estimated as $O(\epsilon)$, which is too large, 
if $(q_{1}-q_{2})$ is not zero \cite{Gedalia:2009kh}.
In order to avoid such too large flavor changing couplings, 
we have to control the coefficients, $ y_{ij} ^{u,d}$, 
and choose appropriate charge assignments. On the other hand, 
large mixing would be required in $(t,\,\! u)$-element, when we discuss the
top forward-backward asymmetry, so that we may have to consider 
non-trivial charge assignments and $ y_{ij} ^{u,d}$ dependence 
for the partially large mixing.\footnote{
Eq. (\ref{eq;mass-FN}) and Eq. (\ref{eq;mix-FN}) lead small mixing 
in $(t,\,\! u)$-element because of the mass hierarchy, $m_u/m_t$.}  

We can also have $\Phi$-quark-quark couplings induced by Yukawa couplings,
\begin{equation}
\label{eq:scalar}%
Y^u_{ij} n^u _{ij}\left(   \frac{\langle H \rangle}{ \langle \Phi \rangle} 
\right)  
\delta \Phi   \overline{U_{Li}}  U_{Rj}   +Y_{ij}^d n^d _{ij}
\left(   \frac{\langle H \rangle}{ \langle \Phi \rangle} \right)  \delta \Phi  
\overline{D_{Li}}  D_{Rj} +h.c.,
\end{equation}
where $\delta \Phi$ is the fluctuation around $\langle \Phi \rangle$.
After changing the base, we could find the flavor changing Yukawa couplings 
because of the charge dependence. 
If the mixing is small, the rotated $Y^u_{ij} n^u _{ij}$ 
and $Y^d_{ij} n^d _{ij}$ would be the same order as $Y^u_{ij}$ and $Y^d_{ij}$, 
and those terms have the suppression of 
$ \langle H \rangle /\langle \Phi \rangle$. 
If $\delta \Phi$ is very light, 
sizable contributions to FCNC could appear, 
but the Yukawa couplings of $\delta \Phi$ are also 
suppressed by $Y^{u,d}_{ij}$.

Eventually, it might be difficult to assign flavor-dependent charge 
according to the realistic Yukawa texture,
because of the FCNC constraints and $A_{\textrm{FB}}$. 
Especially the flavor-dependent charge 
assignment to the down-type quarks would generally cause the problem 
without controlling the coupling constants, $ y_{ij} ^{d}$, 
because of the sizable mixing for CKM matrix.
In the next subsection, we consider models with universal charges 
in the down-type quarks, which can be expected to evade the strong bounds 
on $B$ and $K$ mixing, and several Higgs fields charged under $U(1)'$.
$\Phi$ is also required to make charged and pseudo-scalar Higgs massive.
Such $U(1)'$ charged Higgs can contribute to $A_{\textrm{FB}}$,
and we discuss $U(1)'$ gauge boson, (pseudo) scalar, and charged 
Higgs interactions in the Sec.\ref{sec:phenomenology}. 
We will show the case with only gauge boson and neutral scalar Higgs boson,
which could be interpreted as the result of 
this FN-type model
\footnote{ In the FN-type model, non-renormalizable operators 
induce the effective Yukawa coupling like the FN model, but we will discuss 
the top FB asymmetry supposing that the charge assignment is fixed 
independent of the Yukawa texture allowing to tune the coupling constants, 
$ y_{ij} ^{u,d}$.  }. 
In fact, we will conclude that it is difficult to enhance $A_{\textrm{FB}}$, 
and avoid the strong bound from the same sign top production in the 
FN-type model.

\section{Simple examples with flavored multi-Higgs doublets 
with nonzero $\bm{U(1)'}$ charges}
\label{sec:multi-Higgs-exam}

In this section, we consider two simple choices of $U(1)^{'}$ charges for the 
RH up-type quarks, 
\[
( u_i ) = (0,0,1)~~~{\rm or}~~~(-1,0,1).
\]
Since the $U(1)'$ is chiral, one has to introduce $U(1)'$-charged Higgs doublets
in order to write down the renormalizable Yukawa interactions, as explained 
in Sec.~\ref{sec:multi-Higgs}. Here we construct the  two-Higgs and 
three-Higgs doublet models corresponding to the above two charge assignments, and work out  
the renormalizable Yukawa interactions explicitly. 
Some of the Yukawa couplings involving the neutral (pseudo) scalar and charged 
Higgs bosons will be flavor changing in the up-type quark sector, and will be used 
in Sec.~\ref{sec:phenomenology} for studying the top FB asymmetry, 
the same sign top pair productions and the $Wjj$ excess at hadron colliders.  

\subsection{Two-Higgs doublet model (2HDM) with $\bm{ (u_i) = (0,\,\! 0,\,\! 1)}$}
\label{sec:2-Higgs}
First, let us consider the model with $(u_1,\,\! u_2,\,\! u_3)=(0,\,\! 0,\,\! 1)$ 
and one extra Higgs doublet $H_3$ that is charged under $U(1)'$:       
\begin{center}
\begin{tabular}{|c|c|c|c|c|c|c|}\hline
       & $SU(3)_c$ & $SU(2)_L$  & $U(1)_Y$ & $U(1)'$      \\ \hline 
$H$    & $1$       & $2$        & $1/2$    & $0$          \\ \hline  
$H_3$  & $1$       & $2$        & $1/2$    & $1$          \\ \hline 
$\Phi$ & $1$       & $1$        & $0$      & $q_{\Phi}$   \\ \hline      
\end{tabular}
\end{center}
$\Phi$ not only breaks $U(1)'$ spontaneously and generates the $Z'$ mass,
but also is required to generate masses of charged and pseudo-scalar Higgs 
fields for  any $U(1)'$ charge assignments. Therefore its $U(1)'$ charge 
($q_{\Phi}$) will be  fixed by $u_i$ (see Eq. (\ref{eq:charged-mass-2HDM})).
Under this assignment, the renormalizable Yukawa couplings are 
written as follows,
\begin{eqnarray}
V_y&=& y^u_{i1} \overline{Q_i} \widetilde{H} U_{R1} +y^u_{i2} 
\overline{Q_i} \widetilde{H} U_{R2} +y^u_{i3} \overline{Q_i} 
\widetilde{H_3} U_{R3}  \nonumber  \\ 
   &&+y^d_{ij} \overline{Q_i} H D_{Rj} + y^e_{ij} 
\overline{L_i}  H E_{Rj}  + h.c..
\end{eqnarray}
Two Higgs doublets $H_3$ and $H$ include two neutral scalar, 
one pair of charged Higgs and one pseudoscalar Higgs fields. 
Their vevs are defined as 
$(\langle H \rangle,\,\! \langle H_3 \rangle) = (v \cos \beta / \sqrt{2}, 
\,\! v \sin \beta / \sqrt{2})$.
 
In the mass basis of SM fermions and the  Higgs bosons, the Yukawa 
couplings of the lightest neutral scalar Higgs boson  (denoted by $h$)  
with the SM fermions are described by $R_u~(g_R^u)$ and their masses, $m_i^{u,d,l}$: 
\begin{equation}\label{eq:Yukawa-neutral}
V_h =  Y^u_{ij}  \overline{ \Hat{U}_{Li}} \Hat{U}_{Rj} h
+Y^d_{ij} \overline{ \Hat{D}_{Li}} \Hat{D}_{Rj} h+Y^e_{ij} \overline{ \Hat{E}_{Li}} \Hat{E}_{Rj} h+h.c.,    
\end{equation}
where $Y^{u,d}_{ij}$ and $Y^{e}_{ij}$ are defined as
\begin{eqnarray}
Y^u_{ij}&=& \frac{m_i^u \cos \alpha}{v \cos \beta} \cos \alpha_{\Phi}
\delta_{ij}+ \frac{2  m_i^u}{v\sin 2 \beta} (g^u_R)_{ij} \sin (\alpha-\beta)\cos \alpha_{\Phi}, 
\label{eq:Yuij}
\\
Y^d_{ij} &=& \frac{m_i^d \cos \alpha}{v\cos \beta} \cos \alpha_{\Phi} \delta_{ij}, \\
Y^e_{ij} &=& \frac{m_i^l \cos \alpha}{v\cos \beta} \cos \alpha_{\Phi} \delta_{ij},
\end{eqnarray}
where $\alpha$ is the mixing of $2$ neutral scalar Higgs fields, and $\alpha_{\Phi}$ is the mixing between the neutral scalar components of $\Phi$ and 
$(H,\,\! H_3)$. 
Note that 
$(g^u_R)_{ij}=(g^u_R)^*_{ji}= (R_u)_{i3} (R_u)^*_{j3}$ is satisfied 
for our choice of $U(1)'$ charges: $(u_i) = (0,0,1)$. 

The couplings of charged Higgs and pseudo-scalar Higgs can be written 
by finding the orthogonal direction of the Goldstone mode.
The couplings of charged Higgs boson $h^{\pm}$ 
to the SM fermions are described as follows:  
\begin{equation}
\label{eq:Yukawa-charged}%
V_{h^{\pm}}= - Y^{u-}_{ij} \overline{ \Hat{D}_{Li}  }\Hat{U}_{Rj}
h^{-}+Y^{d+}_{ij} \overline{\Hat{U}_{Li}} \Hat{D}_{Rj} h^{+} 
+h.c.,    
\end{equation}
where $Y^{u-}_{ij}$ and $Y^{d+}_{ij}$ are defined as,
\begin{eqnarray}
Y^{u-}_{ij}&=&  \sum_l (V_{\textrm{CKM}})^*_{li} 
\left \{ \frac{\sqrt{2} m_l^u \tan \beta}{v } \delta_{lj}
- \frac{2 \sqrt{2} m_l^u}{v\sin 2 \beta} (g^u_R)_{lj} \right \}, \nonumber \\
Y^{d+}_{ij}&=& (V_{\textrm{CKM}})_{ij} \frac{ \sqrt{2} m_j^d \tan \beta }{v},
\end{eqnarray}
where $V_{\textrm{CKM}}$ is defined as $(L_u)_{il} (L_d)^*_{jl}$.

The couplings of the pseudo-scalar Higgs, $a$, 
are also described as follows: 
\begin{equation}
\label{eq:Yukawa-pseudo}%
V_{a}= -i  Y^{au}_{ij}  \overline{ \Hat{U}_{Li}  }\Hat{U}_{Rj} a
+i Y^{ad}_{ij} \overline{  \Hat{D}_{Li}} \Hat{D}_{Rj} a +i Y^{ae}_{ij} 
\overline{  \Hat{E}_{Li}} \Hat{E}_{Rj} a 
+h.c.,   
\end{equation}
where $Y^{au,d}_{ij}$ and $Y^{ae}_{ij}$ are defined as   
\begin{eqnarray}
\label{eq:Yauij}
Y^{au}_{ij}&=&  \frac{m_i^u \tan \beta}{v } \delta_{ij}
-  \frac{2 m_i^u}{v\sin 2 \beta}  (g^u_R)_{ij}, \\
Y^{ad}_{ij}&=&  \frac{m_i^d \tan \beta}{v } \delta_{ij}, \\
Y^{ae}_{ij}&=&  \frac{m_i^l \tan \beta}{v } \delta_{ij}.
\end{eqnarray}  

In this model, we also have to avoid large $(u,\,\! c)$ elements 
for the bound from $D^0 - \overline{D^0}$ mixing.  
The small $(u,\,\! c)$ gauge coupling, $(g^u_R)_{12}$, would require 
small $(R_u)_{13}$ or $(R_u)_{23}$.
Fortunately, small  $(g^u_R)_{12}$ could guarantee small $(u,\,\! c)$ elements 
of Yukawa couplings in this model, and there would be no trouble
with $D^0 -\overline{D^0}$ mixing.

Finally, the charged Higgs and pseudo-scalar Higgs must be massive.
The origin of the mass terms may be the following terms, 
\begin{equation}
\label{eq:charged-mass-2HDM}%
\mu H_3^{\dagger} H  (\Phi)^{\frac{1}{q_{\Phi}}} +h.c.,
\end{equation}
where $\mu$ is a coupling constant.
If the renormalizability of this term is required, 
$q_{\Phi}$ should be $1$ or $1/2$.

We may consider the same set we introduced in Sec.~\ref{sec:anomaly} 
to cancel the anomaly: one generation and SM gauge vector-like pairs. 
However, we may need another extra Higgs doublet
for the heavy extra quarks, because  $U(1)'$ charges of 
$U^{\prime}_R$ and $H_3$ are $-1$ and $+1$, and forbid the
Yukawa couplings, $\overline{Q'_L} \widetilde{H}U^{\prime}_R$ and $\overline{Q'_L} \widetilde{H_3}U^{\prime}_R$. 
As mentioned in the Ref.~\cite{Ko-Top}, we can define all fermions 
in the extra generation as opposite-chirality fields
with the same charge assignment as the third generation. 
However, such exotic extra generation allows the mass mixing between 
the SM fermions and extra fermions, like $\overline{D'_L}D_{Ri}$, 
where $D_L^\prime$ is the extra left-handed down-type quarks,
so that we assume that such unfavored 
mass mixings  are sufficiently small enough 
in order to avoid large FCNC contributions. 
We could find simpler solutions for for $ (0,\,\! 0,\,\! 1)$ assignment: 
for example, only one SM gauge vector-like pair, whose $U(1)'$ charges are 
$ (Q_L,\,\! Q_R)= (1,\,\! 0)$. Even in this case, mass terms 
between the extra left-handed fermion and SM fermion, 
which cause FCNC problems, would be allowed by the symmetries.

\subsection{Three-Higgs doublet model (3HDM) with 
$\bm{ (u_i) = (-1,\,\! 0,\,\! 1)}$}
\label{sec:3-Higgs}%
Next we consider a model with $(u_1,\,\! u_2,\,\! u_3)=(-1,\,\! 0,\,\! 1)$, 
which requires three Higgs doublets.
This assignment makes Higgs sector more complicated, but the condition 
for anomaly cancellation does not require any extra generation, 
so that we do not need discuss FCNC constraints from, 
for instance, mixing between the SM leptons and extra leptons. 
This would be a nice property of this three-Higgs doublet model.  

For this $U(1)'$ charge assignment, we need to add two more Higgs 
doublets $H_1$ and $H_3$ that are charged under $U(1)'$ in order to 
write renormalizable Yukawa couplings and get realistic mass matrices:  
\begin{center}
\begin{tabular}{|c|c|c|c|c|c|c|}\hline
       & $SU(3)_c$ & $SU(2)_L$  & $U(1)_Y$ & $U(1)'$      \\ \hline  
$H_1$  & $1$       & $2$        & $1/2$    & $-1$         \\ \hline 
$H_2$  & $1$       & $2$        & $1/2$    & $0$          \\ \hline 
$H_3$  & $1$       & $2$        & $1/2$    & $1$          \\ \hline 
$\Phi$ & $1$       & $1$        & $0$      & $q_{\Phi}$   \\ \hline      
\end{tabular}
\end{center}
The Yukawa couplings are as follows,
\begin{eqnarray}
V_y&=& y^u_{i1} \overline{Q_i} \widetilde{H_1} U_{R1} +y^u_{i2} 
\overline{Q_i} \widetilde{H_2} U_{R2} +y^u_{i3} \overline{Q_i} 
\widetilde{H_3} U_{R3}  \nonumber  \\ 
   &&+y^d_{ij} \overline{Q_i} H_2 D_{Rj} + y^e_{ij} \overline{L_i}  H_2 E_{Rj}.  
\end{eqnarray}
After the EWSB and $U(1)'$ breaking, $H_1$, $H_2$ and $H_3$ will contain 
three neutral scalar, two pairs of charged scalar and two pseudoscalar 
Higgs fields. They mix with each other as well as 
with $\Phi$. 
Let us express the interaction eigenstates 
in terms of mass eigenstates: 
\begin{eqnarray}
\widetilde{h_I} &=& O_{IJ}^h h_J ,  
\nonumber  \\ 
\widetilde{h^{\pm}_n} &=& O^c_{nm}  h^{\pm}_m ,
\nonumber  \\
\widetilde{a_n} &=& O^a_{nm}  a_m ,
\end{eqnarray} 
where $\widetilde{h_I}$, $\widetilde{h^{\pm}_n}$ and $\widetilde{a_n}$ are
scalar, charged and pseudo-scalar Higgs fields in the interaction basis, 
respectively, with $I,J=1,\,\! 2,\,\! 3,\,\! \Phi$ and $n,m=1,\,\! 2, \,\! 3$, 
and the fields in the righthand side $h_J, h^\pm_m$ and $a_m$ are in the mass basis, 
and the Goldstone modes are $a_3$ and $h^{\pm}_3$.
The mixing matrices $O^h$, $O^c$ and $O^a$ are $4 \times 4$, 
$3 \times 3$ and  $3 \times 3$ orthogonal matrices, respectively.  

In the mass basis of SM fermions and the lightest Higgs, $h \equiv h_1$, 
the Yukawa couplings, $Y^{u,d}_{ij}$ and  $Y^{e}_{ij}$, which are defined 
in the Eq.~(\ref{eq:Yukawa-neutral}), are described by $R_u$ and $m_i^u$,
\begin{eqnarray}
Y^u_{ij}&=& \sum^3_{k=1} O_{k1}^h \frac{m_i^u}{ \sqrt{2}\langle H_k \rangle} 
(R_u)_{ik} (R_u)^*_{jk},  \\
Y^d_{ij}&=& O_{21}^h \frac{m_i^d}{\sqrt{2} \langle H_2 \rangle} \delta_{ij},
\\
Y^e_{ij}&=& O_{21}^h \frac{m_i^l}{\sqrt{2} \langle H_2 \rangle} \delta_{ij}.
\end{eqnarray}

The couplings, $Y^{u-}_{ij}$ and $Y^{d+}_{ij}$, 
in the Eq.~(\ref{eq:Yukawa-charged}) of the lightest charged Higgs boson, 
$h^{\pm} \equiv h_1^\pm$, are also described by 
\begin{eqnarray}
Y^{u-}_{ij}&=& \sum^3_{k=1} O_{k1}^c 
\sum_l (V_{\textrm{CKM}})^*_{li} \frac{m_l^u}{\langle H_k \rangle} 
(R_u)_{lk} (R_u)^{*}_{jk}, 
\nonumber \\
Y^{d+}_{ij}&=& O_{21}^c (V_{\textrm{CKM}})_{ij} \frac{m_j^d}{\langle H_2 \rangle}.
\end{eqnarray}
The couplings of the lightest pseudo-scalar Higgs, $a \equiv a_1$, 
in the Eq.~(\ref{eq:Yukawa-pseudo}) are also given by  
\begin{eqnarray}
Y^{au}_{ij}&=& \sum^3_{k=1} O_{k1}^a 
\frac{m_i^u}{ \sqrt{2}\langle H_k \rangle} (R_u)_{ik} (R_u)^*_{jk}, 
\nonumber \\
Y^{ad}_{ij}&=& O_{21}^a \frac{m_i^d}{\sqrt{2} \langle H_2 \rangle} \delta_{ij},
\nonumber \\
Y^{ae}_{ij}&=& O_{21}^a \frac{m_i^l}{\sqrt{2} \langle H_2 \rangle} \delta_{ij}.
\end{eqnarray}

In this model, we also have to avoid large $(u,\,\! c)$ elements 
for the bound from $D_0-\overline{D_0}$ mixing.  
The gauge coupling would require small  $(1,\,\! 2)$ and $(2,\,\! 3)$ mixing 
in $(R_u)_{ij}$, such as $(R_u)_{12}$ and $(R_u)_{13}$.
The Yukawa coupling is not so simple, compared with Sec.~\ref{sec:2-Higgs}, 
but we are sure that the small $(1,\,\! 2)$ and $(2,\,\! 3)$ mixing could also 
realize small $(u,\,\! c)$ elements in the Yukawa couplings, 
as we describe in the appendix in detail.

Besides, we do not need the extra generation for anomaly cancellation  
which caused  FCNC from mass mixing for the choice of $(0,0,1)$,  
because of $u_1+u_2+u_3=0$. 
Only SM gauge vector-like pairs, such as $(q_{LI},\,\! q_{RI})$ introduced 
in Sec.~\ref{sec:anomaly},
are required, and $(Q_L, \,\! Q_R)$ must be defined as $(-1/2, \,\! -3/2)$ to cancel $U(1)_Y U(1)'^2$ product.
When $q_{\Phi}$ is set to $1$, the charged Higgs masses and pseudo-scalar masses 
can be given by the following terms,

\begin{equation}
\mu_{12} \Phi H_2^{\dagger} H_1  + \mu_{23} \Phi  H^{\dagger}_3 H_2 
+ h_{13} \Phi^2  H^{\dagger}_3  H_1+h.c..
\end{equation}
If we assume that the mass terms of $H_1$ and $H_3$, $m_{H_1}^2 |H_1|^2$ 
and $m_{H_3}^2 |H_3|^2$, are very large, we can integrate out $H_1$ 
and $H_3$, and we could realize effective one $SU(2)_L$ doublet model 
which corresponds to the FN-type model,
\begin{equation}
V_y= y^u_{i2} \overline{Q_i} \widetilde{H_2} U_{R2}  -  y^u_{i1 } 
\left(\frac{\mu_{12} \Phi}{m^2_{H_1}}\right) 
\overline{ Q_i}\widetilde{H_2}  U_{R1}-  y^u_{i3 } 
\left(\frac{\mu^*_{23} \Phi^{\dagger}}{m^2_{H_3}}\right) 
\overline{ Q_i} \widetilde{H_2}  U_{R3} +\dots +h.c..
\end{equation}
However, in this parameter region, only neutral scalar Higgs boson would 
contribute to the low-energy physics.  In the next section, we will discuss 
a simple case like the FN-type model with only neutral scalar Higgs,
but we will notice that pseudo-scalar and charged Higgs play an important role 
to evade the same-sign top bound.  In such a case, at least two of 
Higgs doublets should have masses around weak scale, with nonzero vevs.
This would be another way to observe that the FN-type model would not
describe the top FB asymmetry and the same sign top pair production 
properly. 

\section{Collider phenomenology}
\label{sec:phenomenology}
In this section, we discuss phenomenology of our models, especially 
the top forward-backward asymmetry, the same sign top pair production
and the $Wjj$ signal. 
In the FN-type model, the Higgs contribution would be generically small, 
because of the small ratio 
$\langle H \rangle/\langle \Phi \rangle$ and the corresponding
Yukawa suppression or assuming that $\Phi$ is very heavy.  
Therefore we can concentrate on the $Z'$ contributions to the processes 
under consideration that are governed by the gauge couplings.
Even if the contribution of the neutral scalar Higgs is dominant, we will find 
that such scenario is not favored by $A_\textrm{FB}$ and 
the same-sign top pair production, as we see 
in Sec.~\ref{sec:FB-Higgs}.
In the multi-Higgs doublet models, we have not only neutral scalar Higgs, 
but also pseudo-scalar and charged Higgs contributions, so that we can 
improve the result of the FN-type model. 
In this section, we focus on the 2HDM or 3HDM.

\subsection{Numerical inputs}
In the numerical calculation, we take the top quark mass $m_t=173$ GeV.
We use CTEQ6m for a parton distribution function~\cite{cteq}.
Both the renormalization and factorization scales are taken to be $m_t$.
We use $K=1.3$ for the $K$ factor in order to take into account the 
QCD radiative correction which is unknown as of now.
The center-of-momentum (CM) energy $\sqrt{s}$ is taken to be
$1.96$ TeV at the Tevatron and $7$ TeV at the LHC, respectively.

In the 2HDM case with the $U(1)^{'}$ charge assignment 
$(u_i)=(0,0,1)$, one can deduce $|(g^u_R)_{ut}|^2 = (g^u_R)_{uu} (g^u_R)_{tt}$
from Eq.~(\ref{eq:gLgR}). 
However in the other multi-Higgs models such as 3HDM with different $U(1)^{'}$ 
charge assignments,  this identity would not be valid in general, 
as can be inferred 
from Eq.~(\ref{eq:gR-3HDM}).  In the numerical analysis, we assign 
$|(g^u_R)_{ut}|^2=(g^u_R)_{uu}(g^u_R)_{tt}$ assuming the 2HDM case. 
In the multi-Higgs models, this assumption might alter the $s$-channel 
$Z^\prime$ contributions to the $t\bar{t}$ production, 
but the deviation would be tiny numerically. 
This is because the contribution to the $t\bar{t}$
production from the diagonal coupling $(g^u_R)_{uu}$ and $(g^u_R)_{tt}$ 
would be small because the interference terms between the SM QCD contribution
at leading-order in $\alpha_s$ and 
the $s$-channel diagram mediated by the $Z^\prime$ boson vanish.

There is no strict bound for the mass $m_{Z^\prime}$ of the $Z^\prime$ boson
in our models, but it is assumed to be around the EW scale.
Since the $Z^\prime$ boson does not couple to the charged lepton and
neutrinos by construction, stringent bounds on an extra $Z^\prime$ boson
from electroweak (EW) precision tests at LEP II and Drell-Yan processes
at the Tevatron and LHC could be avoided.
If the $Z^\prime$ boson is heavier than the top quark mass and
has diagonal couplings to the SM quarks,
its mass could be constrained by dijet production results at hadron colliders.
In the relatively low mass region the UA2 experiments give
strongest bounds for an $s$-channel resonance,
while the Tevatron experiments strongly constrain its mass
in the high mass region of the resonance \cite{Fan:2011vw}. 
Below the region lighter than the top quark the constraint is rather weak.

As discussed before,
much attention on the light $Z^\prime$ boson scenarios with the mass
around 140 $\sim$ 160 GeV has been paid,
motivated by the recent collider and dark matter experiments.
In this work, we consider two cases: $m_{Z^\prime}=145$ GeV and
$m_{Z^\prime}=160$ GeV. Then the top quark can decay into a $Z^\prime$ boson
with an up quark because there is a flavor-changing neutral current
in the $u_R$-$t_R$-$Z^\prime$ vertex. It could alter significantly  
the branching ratio of the top quark to $Wb$. We assume the branching
ratio of the top quark to $Z^\prime u$ is less than 5 \%.
Then the coupling $\alpha_x = (g^\prime (g^u_R)_{ut})^2/(4\pi)$ should be
less than $0.012$ for $m_{Z^\prime}=145$ GeV, but it is not constrained
in the region $\alpha_x \leq 0.025$ for $m_{Z^\prime}=160$ GeV.

In the most general case, 4 Higgs doublets are required to write down
proper mass terms for the SM fermions. As simpler cases we introduced 
2HDM and 3HDM in Secs.~\ref{sec:2-Higgs} and \ref{sec:3-Higgs}.
Thus at least $3$ neutral Higgs bosons and $1$ charged Higgs boson pair exist.
In this work we assume that only the lightest scalar, pseudo-scalar and
charged Higgs bosons are relevant in the $t\bar{t}$ and $Wjj$ production
for simplicity. The Yukawa couplings could be proportional to the fermion
masses after EW and $U(1)^\prime$ breaking. Thus we assume that
the scalar and pseudo-scalar Higgs bosons have large off-diagonal terms
$Y_{tu}(\equiv Y^{u}_{31})$ and 
$Y_{tu}^a(\equiv Y^{au}_{31})$ only for the top and up quarks, 
which may be natural in our model 
(see Eq.s~(\ref{eq:Yuij}), (\ref{eq:Yauij}), (\ref{eq:Yuijm})
and (\ref{eq:chargedm}) ).
The other Higgs bosons are assumed to be sufficiently heavy to have no effects
on the $t\bar{t}$ and $Wjj$ production.

In our model, the top quark may decay into a Higgs boson if Higgs boson 
is lighter than the top quark. If both the $Z^\prime$ and Higgs bosons
are lighter than the top quark, the branching fraction of
the top quark decay to the non-$Wb$ state could easily be over 10 \%,
which might be dangerous. In order to avoid this harmful situation, 
we assume that all the Higgs bosons are heavier than the top quark. 
In this work, as an illustration of our model, 
we take $m_h=180$ GeV, $m_a=300$ GeV and $m_{h^+}=270$ GeV, respectively. 
The mass of the scalar Higgs boson looks like conflict
with the mass bounds from the recent CMS and ATLAS experiments,
which exclude the mass range from 149 GeV to 206 GeV and 
from 155 GeV and 190 GeV at 95 \% C.L., respectively~\cite{CMSHiggs,ATLASHiggs}.
However in our model the mass bound of the lightest Higgs boson should be 
weaker, since new decay channels of Higgs boson, such as 
$h\to t\bar{u}$, $h\to$ $\Phi+$anything etc., will be open.

\begin{figure}[!t]
\begin{center}
\epsfig{file=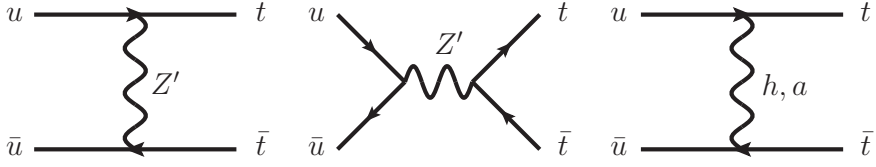,width=0.7\textwidth}
\caption{\label{fig:ttbar}%
The Feynman diagrams for the $t\bar{t}$ production.
}
\end{center}
\end{figure}

\begin{figure}[!t]
\begin{center}
\epsfig{file=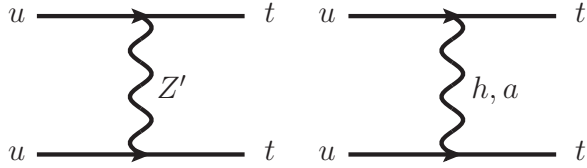,width=0.467\textwidth}
\caption{\label{fig:tt}%
The Feynman diagrams for the $t t$ production.
}
\end{center}
\end{figure}

\subsection{Top physics}
\subsubsection{Empirical data}
\label{sec:empirical}
We show the Feynman diagrams involving new physics contributions 
to the $t\bar{t}$ production 
in Fig.~\ref{fig:ttbar}.
The $Z^\prime$ boson contribute to the $t\bar{t}$ production 
through its $t$- and $s$-channel exchange in the parton process
$u\bar{u}\to t\bar{t}$, while the $h$ and $a$ bosons contribute
only through a $t$-channel diagram.
So far the $t\bar{t}$ pair production cross section is in good agreement with
the SM predictions at the Tevatron and LHC. The empirical cross sections are
$\sigma(t\bar{t}) = ( 7.5 \pm 0.48 )$ pb at the Tevatron~\cite{cdfttbar}
and $\sigma(t\bar{t}) = ( 158 \pm 19 )$ pb at CMS~\cite{CMSttbar}, 
respectively. In this work we use the Tevatron result in order to check 
our model, which is more sensitive to the $u\bar{u}\to t\bar{t}$ process.

The top forward-backward asymmetry $A_{\textrm{FB}}$ in the $t\bar{t}$ 
rest frame is defined by the difference of the top quark numbers
in the forward region and in the backward region. The SM prediction 
at next-to-leading order (NLO) is
$A_\textrm{FB}=0.058\pm 0.009$ at the Tevatron~\cite{CDFAFB1}. 
The CDF Collaboration reported 
$A_{\textrm{FB}}^{\textrm{lepton+jets}}= (0.158 \pm 0.075)$
in the lepton+jets channel with an integrated luminosity 
of $5.3$ fb$^{-1}$~\cite{CDFAFB1}
and $A_{\textrm{FB}}^{\textrm{dilepton}}=0.42 \pm 0.17$ 
in the dilepton channel~\cite{CDFAFB2}. In the dilepton channel
the central value of the asymmetry is quite large, but its uncertainty
is also large. Both are consistent with each other within $1.5$ $\sigma$ level.
A similar deviation in the top forward-backward asymmetry has recently
confirmed by the D0 Collaboration with
$A_{\textrm{FB}}=0.196\pm 0.065$ in the lepton+jets channel
with an integrated luminosity of $5.4$ fb$^{-1}$~\cite{D0AFB}.
For illustration we use the result in the lepton+jets channel at CDF.

In our model the $t$-channel diagrams play a key role in accommodating
all the empirical data. 
It is known that models with a light $Z^\prime$ boson or a light scalar boson 
with FCNC are strongly constrained by 
the same sign top pair production at the LHC ~\cite{Cao:2011ew,Berger:2011ua}. 
Up to the present the most stringent bound for the same sign top quark
pair production is given by the CMS Collaboration: 
$\sigma(tt) < 17$ pb~\cite{CMSsametop}.

\begin{figure}[!t]
\begin{center}
\epsfig{file=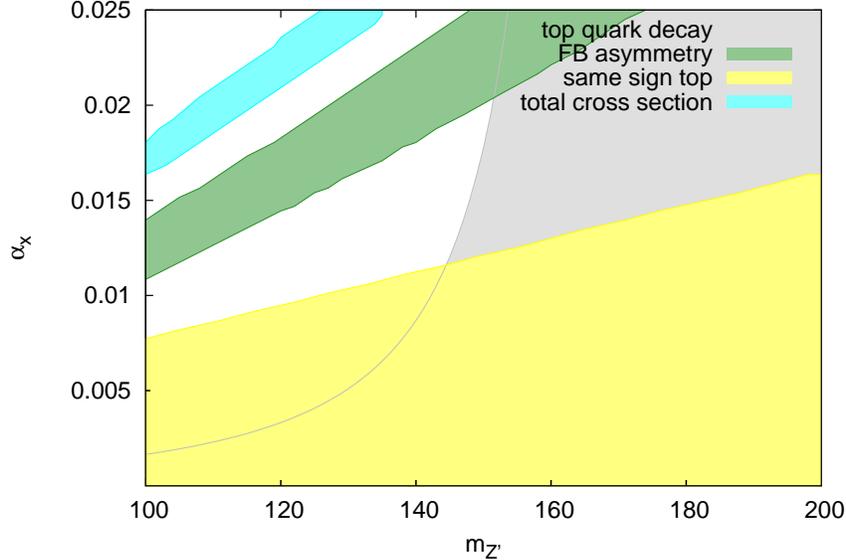,width=0.7\textwidth}
\vspace{-3ex}
\caption{\label{fig:axMzp}%
The favored region for $\alpha_x$ and $m_{Z'}$ in the case where 
only $Z^\prime$ contributes.
}
\end{center}
\end{figure}

\subsubsection{$Z^\prime$-dominant case}
\label{sec:Zp}
In this section, we consider the case that only the $Z^\prime$ boson
contribute to the $t\bar{t}$ production at the Tevatron, 
assuming that $(g_R^u)_{ut}$, $(g_R^u)_{uu}$, and $(g_R^u)_{tt}$ are dominant
over other elements of the coupling matrix $(g_R^u)_{ij}$.
This is similar to the simple phenomenological 
model suggested by Jung, Murayama, Pierce and Wells~\cite{Jung:2009jz}.
One difference is that the $s$-channel contribution of $Z'$ was ignored in 
Ref.~\cite{Jung:2009jz}. 

Figure \ref{fig:axMzp} represents the allowed region 
for $m_{Z^\prime}$ and $\alpha_x$, which is in agreement with each experiment.
The cyan band corresponds to the region satisfying the $t\bar{t}$ 
production cross section at the Tevatron in the 1-$\sigma$ level,
while the green band satisfies the top forward-backward asymmetry at CDF
in the 1-$\sigma$ level, respectively. The yellow region represents
the region in which the same sign top pair production at the LHC
is less than the upper bound at CMS. The branching fraction of the top quark
to $Z^\prime u$ is less than 5 \% in the gray region. There is no overlapped 
region from the four constraints we consider. In particular it is quite difficult 
to evade the strong constraint from the same sign top pair production at the LHC.
Figure~\ref{fig:axMzp} tells us that the region consistent with Tevatron and LHC 
could not realize the large $A_\textrm{FB}$ which CDF and D0 observe.
Thus a simple $Z^\prime$-exchange model with a large FCNC is excluded.

\begin{figure}[!t]
\begin{center}
\epsfig{file=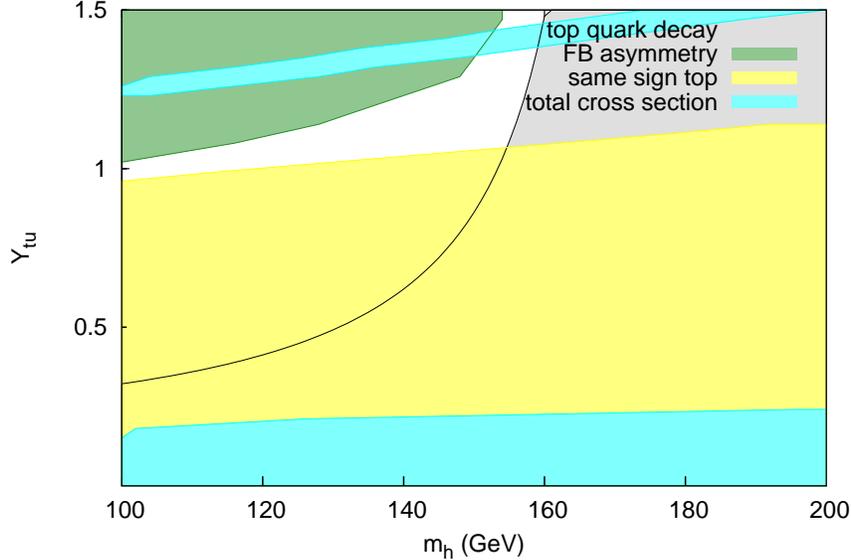,width=0.7\textwidth}
\vspace{-3ex}
\caption{
The favored region for $Y_{tu}$ and $m_h$ in the case where only $h$ 
contributes.
\label{fig:scalar}%
}
\end{center}
\end{figure}

\begin{figure}[!t]
\begin{center}
\epsfig{file=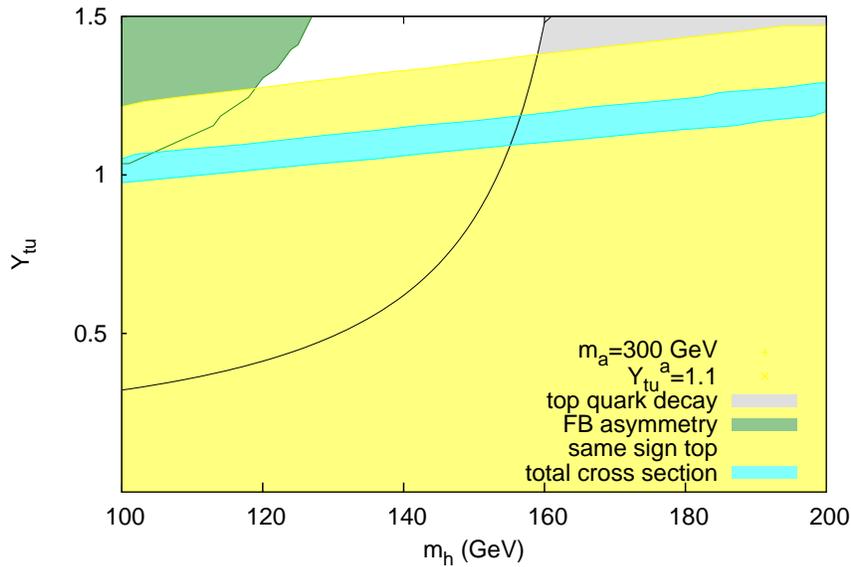,width=0.7\textwidth}
\vspace{-3ex}
\caption{
The favored region for $Y_{tu}$ and $m_h$ in the case where $h$ and $a$
contributes.
\label{fig:pseudoscalar}%
}
\end{center}
\end{figure}

\subsubsection{Higgs-dominant case}
\label{sec:FB-Higgs}
In this subsection, we discuss top forward-backward asymmetry 
and the constraints from Tevatron and LHC in the multi-Higgs models. 
As discussed in the Ref.~\cite{Babu, Blum:2011fa}, 
the Higgs contributions can enhance $A_\textrm{FB}$.
According to Sec.~\ref{sec:multi-Higgs},
we have not only $Z'$ contribution but also neutral (pseudo) scalar, 
and charged Higgs contribution. In this section we consider only
the Higgs contribution by assuming that the gauge coupling of the $Z^\prime$ 
boson is negligible. The scalar and pseudo-scalar Higgs bosons contribute
to the $t\bar{t}$ and $tt$ production only through the $t$-channel diagrams.

If the mixing, especially $(R_u)_{13}$, is large, 
the $(t,\,\! u)$ element of Yukawa coupling for neutral scalar 
and pseudo-scalar Higgs could be enhanced.
Since the Higgs masses are determined by their own vevs,
they cannot be arbitrarily heavy, and we could expect them 
(at least the lightest one) to contribute to the top physics.
$h$ and $a$ bosons can contribute to the same sign top pair production 
through their $t$-channel exchanges.
In order to avoid the FCNC contribution of Higgs exchanges 
\cite{Blum:2011fa}, we adopt the parametrization which 
we discussed in the subsections \ref{sec:2-Higgs}, \ref{sec:3-Higgs}, 
and the appendix.

We first consider the case without pseudo-scalar Higgs boson by 
setting $Y_{tu}^a$ to zero.  In Fig.~\ref{fig:scalar} 
we show the each region allowed by the $t\bar{t}$ production (cyan), 
the same-sign top (yellow), the width of $t$ (gray), and $A_{\rm FB}$ (green), 
based on the bounds we discussed in the \ref{sec:empirical}.
In order to satisfy the bound from the same sign top pair production,
$Y_{tu}$ should be less than 1, but it is impossible to get enough 
enhancement for $A_\textrm{FB}$. Thus a simple model with a scalar Higgs boson
is excluded.

Next, we consider the case where both $h$ and $a$ contribute to 
the $t\bar{t}$ and $tt$ production through their $t$-channel exchanges.
For simplicity we assume the Yukawa coupling $Y_{tu}^a=1.1$ and the mass
$m_a=300$ GeV of the pseudo-scalar Higgs $a$.
We show the allowed region by the $t\bar{t}$ production (cyan),
the same-sign top (yellow), the width of $t$ (gray), and $A_{\rm FB}$ (green),
respectively, in Fig.~\ref{fig:pseudoscalar}.
In the low mass region $m_h\sim 100$ GeV,
there is an overlap region which satisfies constraints from
the $t\bar{t}$ production, $tt$ production rates and $A_\textrm{FB}$,
but the branching fraction of the top quark decay to the Higgs boson becomes  
quite large. 
Therefore the model with scalar and pseudo-scalar Higgs bosons is excluded.

\begin{figure}[!t]
\begin{center}
\epsfig{file=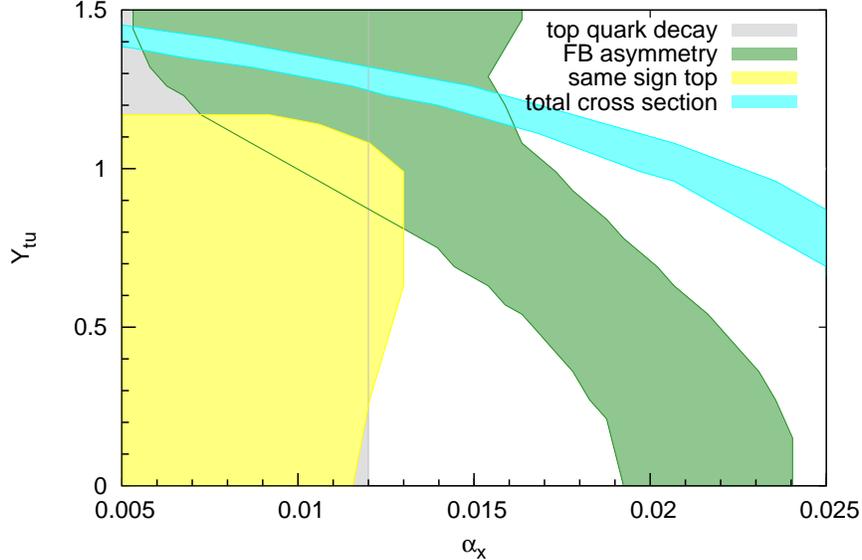,width=0.7\textwidth}
\vspace{-3ex}
\caption{
The favored region for $\alpha_x$ and $Y_{tu}$ for
$m_{Z^{\prime}}=145$ GeV and $m_h=180$ GeV.
\label{fig:zh}%
}
\end{center}
\end{figure}

\begin{figure}[!t]
\begin{center}
\epsfig{file=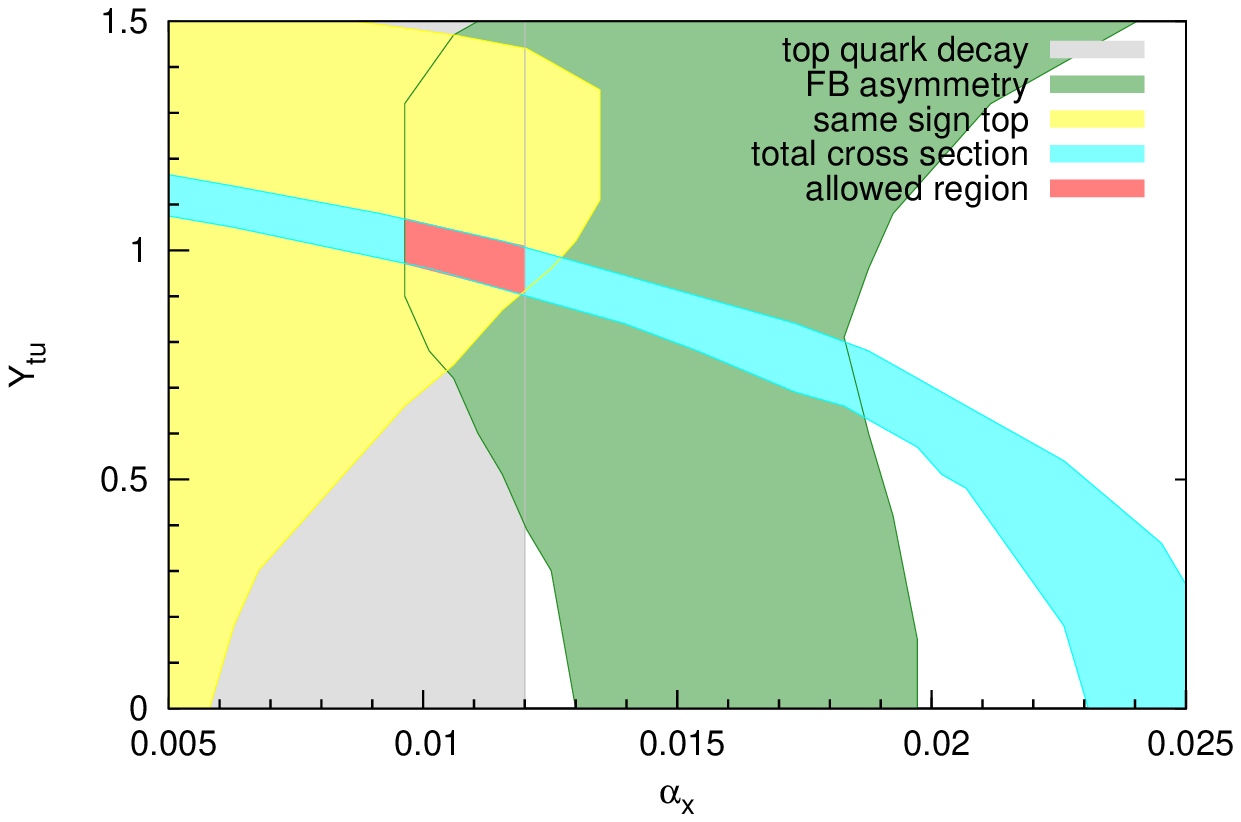,width=0.7\textwidth}
\vspace{-3ex}
\caption{
The favored region for $\alpha_x$ and $Y_{tu}$ 
at $(m_{Z^{\prime}},\,\! m_h,\,\! m_a )=(145,\,\! 180,\,\! 300)$ GeV 
and $Y_{tu}^a=1.1$.
\label{fig:axYtu-145}%
}
\end{center}
\end{figure}

\begin{figure}[!t]
\begin{center}
\epsfig{file=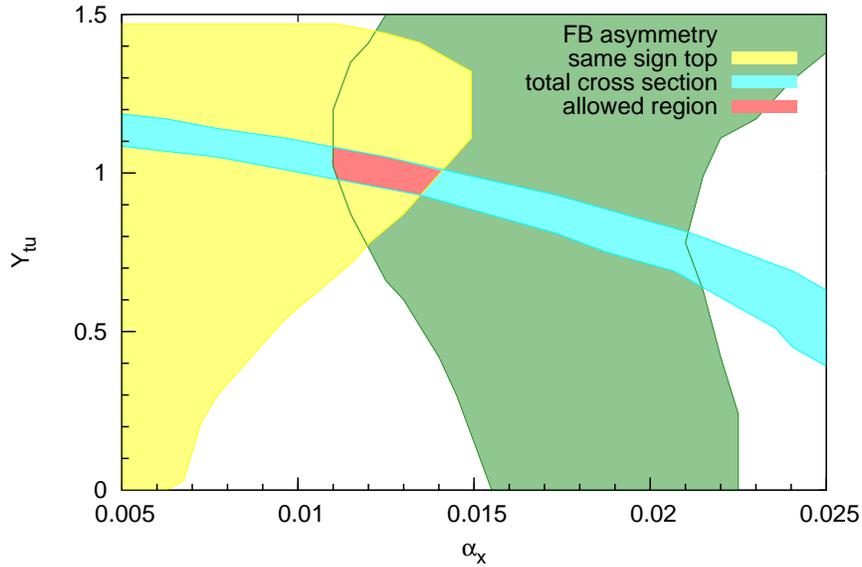,width=0.7\textwidth}
\vspace{-3ex}
\caption{
The favored region for $\alpha_x$ and $Y_{tu}$ 
at $(m_{Z^{\prime}},\,\! m_h,\,\! m_a )=(160,\,\! 180,\,\! 300)$ GeV 
and $Y_{tu}^a=1.1$.
\label{fig:axYtu-160}%
}
\end{center}
\end{figure}

\subsubsection{General case}
\label{sec:mixed}
In this subsection we consider more general case where $Z^\prime$, $h$ and $a$
contributes to the $t\bar{t}$ and $tt$ production.
We assume that the $Z^\prime$ boson has only one large off-diagonal 
element $({g^u_R})_{ut}$ and two large diagonal elements $({g^u_R})_{uu}$ 
and $({g^u_R})_{tt}$ as in the previous case\footnote{In the 2HDM, 
$|(g^u_R)_{ut} |^2=(g^u_R)_{uu}(g^u_R)_{tt}$ is satisfied, 
and in the 3HDM, $(g^u_R)_{uu}$ and $(g^u_R)_{tt}$ could be smaller 
than $(g^u_R)_{ut}$, as seen in Eq. \ref{eq:gR-3HDM}.}. 
As we see in the subsections \ref{sec:2-Higgs} and \ref{sec:3-Higgs}, 
and the appendix, large $(g^u_R)_{ut}$ could enhance the $Y_{tu}$ 
and $Y_{tu}^{a}$ elements in 2HDM and 3HDM,
as far as $H_2$ is not a dominant component in the lightest Higgs 
boson $h$. 
On the other hand, small $(g^u_R)_{cc}$, $(g^u_R)_{uc}$,  
and $(g^u_R)_{ct}$ could realize small $(u,\,\! c)$ element
of the Yukawa couplings for neutral scalar and pseudo-scalar Higgs bosons, 
which gives the small $D^0$-$\overline{D^0}$ mixing.  
If either $\tan \beta$ or $\tan \beta_{1,2}$ are large, 
$Y_{ut}$ and $Y^{a}_{ut}$ may be also enhanced.  
Also the  Yukawa coupling for the charged Higgs boson 
$Y^{u-}_{dt}$ and $Y^{u-}_{bt}$, which contribute to the 
$B_d$-$\overline{B}_d$ mixing at one-loop level, could also be sizable.
In order to avoid this situation, we will concentrate on the small 
$\tan \beta$ region in the following.  Then, only the Yukawa couplings 
for the neutral scalar and the pseudo-scalar bosons,   
$Y_{tu}$, $Y_{tt}$, $Y_{tu}^a$, and $Y^a_{tt}$, could be $\sim O(1)$. 
In the charged Higgs sector,  only $Y^{u-}_{bu}$ and  $Y^{u-}_{bt}$ are 
dominant in the small $\tan \beta$ case,  which may not contribute 
significantly to the top physics we are interested in this paper.

First, we consider the case where only $Z^\prime$ and $h$ 
contribute to the $t\bar{t}$ and $tt$ production by setting $Y_{tu}^a$ to zero.
In Fig.~\ref{fig:zh}, we show the each region allowed 
by the $t\bar{t}$ production (cyan), the same-sign top (yellow), 
the width of $t$ (gray), and $A_{\rm FB}$ (green)
for $m_{Z^\prime}=145$ GeV and $m_h=180$ GeV, respectively.
As seen in Fig.~\ref{fig:zh}, there is no region which satisfies all the 
constraints from the collider experiments. In order to be consistent with
the upper bound for the same sign top pair production at CMS,
the gauge coupling $\alpha_x$ and Yukawa coupling $Y_{tu}$ should be smaller 
than $0.13$ and $1.2$, but in that region the $t\bar{t}$ cross section
at the Tevatron turns out to underestimate the empirical data.
As we mentioned before, the FN-type model could have the neutral scalar Higgs 
contribution through the mixing between $\Phi$ and $H$, but it is impossible 
to enhance $A_{\rm FB}$ enough because of the same-sign top bound.

However in the multi-Higgs models, we could have another contribution 
such as pseudo-scalar Higgs, $a$.
As we see in the Eq.~(\ref{eq:Yukawa-neutral}) and (\ref{eq:Yukawa-pseudo}), 
$h$ and $a$ couplings have opposite signs for the same-sign top pair production,
leading to the destructive interference.  
Therefore we could expect that the interference among $a$, $h$, 
and $Z^{\prime}$ decrease the $tt$ cross section.
For simplicity we assume $Y_{tu}^a=1.1$ and $m_a=300$ GeV.
Figures~\ref{fig:axYtu-145} and \ref{fig:axYtu-160} show the 
allowed regions corresponding to the each experimental bounds in the cases 
with $m_{Z^{\prime}}=145$ GeV and $m_{Z^{\prime}}=160$ GeV.
We see that the light $Z^{'}$ scenario survives the same
sign top pair constraint due to the destructive interference from $h$ and 
$a$ contributions in the $t$-channel.
The $160$ GeV $Z^{\prime}$ can drastically reduce the branching fraction 
of $t \rightarrow Z^{\prime}u$, so that
the gray region need not be included in the Fig. \ref{fig:axYtu-160}.
The red region is consistent with all 
the experimental results from the Tevatron and LHC up to now. 
As one can check explicitly using Eq.s (\ref{eq:Yuij}) and (\ref{eq:Yauij}), 
$Y_{tu}$ must be smaller than $Y^a_{tu}$ in the 2HDM.
The allowed regions in the Figs.~\ref{fig:axYtu-145} and \ref{fig:axYtu-160}
satisfy this condition.
Here we ignore the  $\sim 5 \%$ asymmetry from the NLO contribution
in the SM.  Adding this to our predictions will make the $A_{\rm FB}$ larger.
$A_\textrm{FB}^\textrm{New}$ in the allowed region in Figs.~\ref{fig:axYtu-145}
and \ref{fig:axYtu-160} is between $0.084$ and $0.12$ 
without the contribution from the SM NLO.
If one includes the contribution from the SM NLO effects, 
$A_\textrm{FB}$ could be enhanced to about 14 \%.

\begin{figure}[!t]
\begin{center}
\epsfig{file=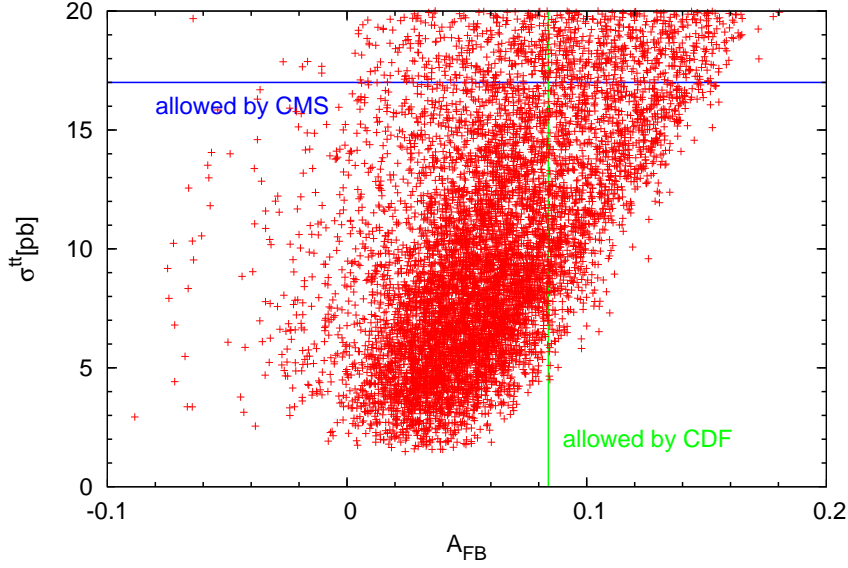,width=0.7\textwidth}
\vspace{-3ex}
\caption{
The scattered plot for the top forward-backward asymmetry at the Tevatron and 
the same sign top pair production at the LHC in unit of pb 
for the model parameters which agree with the cross section for
the $t\bar{t}$ pair production at the Tevatron.
\label{fig:AFBtt}%
}
\end{center}
\end{figure}

Now we examine our model by varying the model parameters
in order to watch how the allowed region is changed.
As we have mentioned earlier, we do not consider the lighter Higgs boson
than the top quark mass to avoid the large exotic decay of the top quark.
$m_{Z^\prime}=145$ GeV is fixed in this analysis.
For the other parameters, we choose the following ranges:
$180 \textrm{ GeV} < m_h, m_a < 1 \textrm{ TeV}$,
$0.005  < \alpha_x < 0.025 $, and
$0.5  < Y_{tu}, Y_{tu}^a < 1.5 $. We impose the condition 
$|Y_{tu}| < |Y_{tu}^a|$ which can be induced from Eq.s~(\ref{eq:Yuij})
and (\ref{eq:Yauij}).
In Fig. \ref{fig:AFBtt}, we show the scattered plot for the top FB
asymmetry at the Tevatron and the same sign top pair production at the LHC
for the above parameter region. All points in the figure satisfy the top quark pair 
$(t\bar{t})$ production rate at the Tevatron within $1\sigma$. The right side 
of the vertical line is consistent with the top FB asymmetry 
in the lepton+jets channel at CDF, while the lower region of the horizontal 
line is the allowed region from the same sign top pair production at CMS.
Therefore the points in the right-lower side are favored by the present 
experiments. The points in the favored region satisfy the following region:
$180 \textrm{ GeV} < m_h < 250 \textrm{ GeV}$, $0.005 < \alpha_x < 0.014$,
$0.75  < Y_{tu} < 1.3 $, and $0.9  < Y_{tu}^a < 1.5 $ with a constraint 
$|Y_{tu}| < |Y_{tu}^a|$.
The lightest scalar Higgs boson is less than 250 GeV to be accommodated with
the data, but the mass $m_a$ of the lightest pseudoscalar Higgs boson 
is not constrained much by the data. In general the cross section 
for the same sign top pair production becomes small as the gauge or Yukawa
coupling becomes small, but there is no general tendency in the dependence on 
the Higgs boson masses.

\begin{figure}[!t]
\begin{center}
\epsfig{file=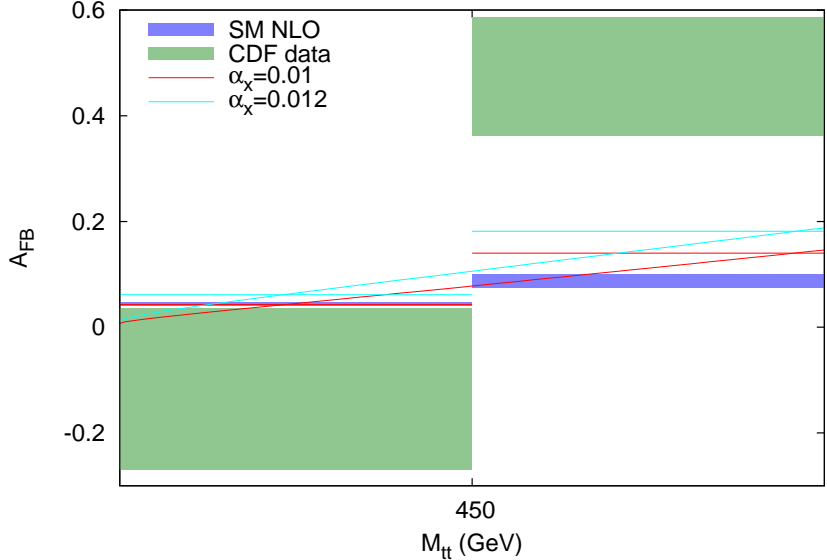,width=0.7\textwidth}
\vspace{-3ex}
\caption{
The top forward-backward asymmetry as a function of the $t\bar{t}$ 
invariant mass at the Tevatron for $m_{Z^\prime}=145$ GeV,
$m_h=180$ GeV, $m_a=300$ GeV and $Y_{tu}^a=1.1$. 
The red and cyan lines correspond to $\alpha_x=0.01$ and
$\alpha_x=0.012$, respectively. 
The blue bins are the SM prediction from {\sc mc@nlo} while
the green bins are measurements in the lepton+jets channel at CDF.
\label{fig:asym}%
}
\end{center}
\end{figure}

The CDF Collaboration also announced the top forward-backward asymmetry
in the lepton+jets channel by dividing the phase space. 
One of the most interesting results is that
$A_{\textrm{FB}}$ in the $t\bar{t}$ invariant mass region larger than 450 GeV
is over about $3.4$ $\sigma$ from the SM prediction.
In Fig.~\ref{fig:asym}, we depict our results for the invariant mass
distribution of $A_{\textrm{FB}}$ for two reference parameter sets:
$\alpha_x=0.01$ (red line or red bin) and
$\alpha_x=0.012$ (cyan line or cyan bin) for $m_{Z^\prime}=145$ GeV,
$m_h=180$ GeV, $m_a=300$ GeV, $Y_{tu}=1$ and $Y_{tu}^a=1.1$.
The green and blue bands correspond to the CDF data in the lepton+jets 
channel and the SM prediction from {\sc mc@nlo}, respectively.
Our prediction in the large $t\bar{t}$ invariant mass region is rather smaller
that the CDF data. However if we include the NLO corrections in the SM
the prediction would agree with the data within about $2$ $\sigma$.

\subsection{Dijet resonance}
Another interesting observation that might be related with light $Z^{'}$ is 
the CDF $Wjj$ excess \cite{CDFWjj}. 
One possible interpretation of which is $p\bar{p} \rightarrow WZ^{'}$ 
followed by $Z^{'} \rightarrow jj$ with $\sigma (WZ^{'} ) \sim 4$ pb
and $m_{Z^{'}} \sim 140$ GeV \cite{Ko:2011ns,Buckley:2011vc,Cheung:2011zt}.  
This excess however was not confirmed by the D0 Collaboration
\cite{D0Wjj}, and more investigation is necessary for understanding 
this discrepancy.  Further data from the LHC for this channel will shed 
light on this issue.

In our model, $Z^{\prime}$ and charged Higgs may play an important role 
in the $p\overline{p} \rightarrow Wjj$ process.
If charges of $U(1)'$ are assigned to quarks universally, gauge couplings 
and Yukawa couplings are flavor-blind,
but we could have large diagonal elements which would be constrained by the UA2 experiment~\cite{ua2,ua2-2}.
The main process would be 
$u,\overline{d}(d,\overline{u}) \rightarrow W^+(W^-), Z^{\prime}$. 
As investigated well in Ref.\cite{Ko:2011ns,Buckley:2011vc,Cheung:2011zt}, 
that cross section could be large to explain the data  without 
conflict with the UA2 bound.

In the case that charges of $U(1)'$ are flavor-dependent,
we may allow only large off-diagonal elements, such as $Y_{tu}$ and 
$\alpha_x$ to enhance the $A_{\rm FB}$.    
In this case, the most important one is the parton process
$u\bar{b}(b\bar{u}) \to h^\pm \to W^\pm Z^\prime$ with a subsequent decay 
$Z^\prime\to j j$,  where $h^\pm$ is the lightest charged Higgs boson.
The charged Higgs boson has a similar coupling structure 
to the neutral coupling, 
so that only the $u_R$-$b_L$-$h^+$ and  $b_R$-$u_L$-$h^-$ vertices 
can be as large as that for the $u_R$-$t_L$-$h$. 
In the 2HDM, the interaction lagrangian for the charged Higgs boson 
with the $W$ and $Z^\prime$ boson is given by
\begin{equation}
\mathcal{L} = - g^\prime m_W \sin 2\beta h^+ {W^-}^\mu Z^\prime_\mu 
+ h.c..
\end{equation}
For  $m_{h^\pm}=270$ GeV, we get $\sigma (W Z^\prime) \sim 10$ pb 
$\times \sin^2 2 \beta \lesssim 10$ pb at the Tevatron. 
It would be about $4.5$ pb for $\sin 2 \beta =0.7$ which is in the range 
of the CDF report,  but could be substantially smaller if  $\sin 2\beta$ 
becomes smaller.
In the 3HDM, $\sin 2\beta$ can be replaced by $\sin 2 \beta_2$ 
for $\xi^{\pm}_1$ (See the appendix).

\subsection{Single top production}
\label{sec:single}
The large flavor changing neutral current in the top sector 
implies a large single top quark production at hadron colliders.
For example, there may be large single top production through
the $g u\to t Z^\prime$ or $t h$ processes in our model.
The single top quark production was measured by the D0 Collaboration
with $\sigma(p\bar{p}\to t b q+X)=2.90\pm 0.59$ pb \cite{D0singletop}.
The CMS Collaboration also announced the cross section for the single top
quark production: 
$\sigma(p\bar{p}\to t b q+X)=83.6\pm 29.8 \pm 3.3$ pb \cite{CMSsingletop}.
The experimental results are based on the observation of a top quark
with an extra $b$ quark, {\it i.e.} require tagging two $b$ quarks.
However in our model
the branching fractions of $Z^\prime$ and $h$ decays to the $b q + X$ 
state are quite small. Thus our model would not be constrained by
the current experiments on the single top production.
Eventually our model would be strongly constrained
if the cross section in the $p \bar{p} (p) \to t + X$ channel is measured.
One possible mode is a $jj$ resonance associate 
with a single top quark~\cite{Jung:2011id}.

\section{Cold dark matter}
\label{sec:CDM}

As discussed in several 
papers~\cite{Ko,Buckley:2010ve,Gondolo:2011eq,Wise,Dulaney,Alwall:2010jc,Feng},
the extension to $U(1)'$ can also provide CDM candidates. 
As we discuss in the subsection \ref{sec:anomaly}, we may require SM gauge 
vector-like pairs for anomaly condition. 
Especially, 
the 3HDM required only $(q_{LI},\,\! q_{RI})$ for the $U(1)_Y U(1)'^2$ anomaly cancellation.
In this section, we comment on the possibility that the required chiral fields 
give rise to CDM candidates.

First, let us focus on models with $SU(2)_L$ doublet pairs, 
$(l_{Ri},\,\! l_{Li})$ instead of $SU(3)_c$ triplets $(q_{LI},\,\! q_{RI})$, 
which was introduced in the subsection \ref{sec:anomaly}.
$l_{Ri}$ and $l_{Li}$  are chiral fields charged under $SU(2)_L \times U(1)_Y$
like the left-handed lepton,  so we can expect one component to be 
charged and the other to be neutral after EW breaking.
The Yukawa couplings corresponding to the mass terms are 
as follows,\footnote{When $q_{\Phi}$ is to $1$, mass terms 
of$(l_{LI},\,\! l_{RI})$ or $(q_{LI},\,\! q_{RI})$ can be given by this form, 
because the solution with $Q_L-Q_R=1$ is always found.  }
\begin{equation}
V_l=   y^l_{1} \Phi^{\dagger} \overline{l_{R1}} l_{L1} + y^l_{2} \Phi 
\overline{l_{R2}} l_{L2}.  
\end{equation}
The charged and neutral fields are degenerate after $U(1)'$ breaking, and
then they can be split after EW breaking according to radiative corrections. 
$l_{Li}$ and $l_{Ri}$ do not have Yukawa coupling with the SM fermions 
because of $U(1)'$, so that $U(1)$ global symmetry, which is phase rotation 
of $l_{Li}$ and $l_{Ri}$, could be remained, and their stability is 
guaranteed by the global symmetry. 
After EWSB, one component is charged and the other is neutral.
The light neutral components could be good CDM candidates, 
as discussed in the Ref. \cite{Cirelli:2005uq}.
They are degenerate before EWSB, but especially the charged particles
get enough large radiative corrections and then heavier than the neutral. 
If the mass split is bigger than the electron mass, the charged can decay 
to the neutral, $e$, and $\nu_e$ through the weak interaction.

The CDM can annihilate to 2 light quarks through $Z'$ and $Z $ boson 
exchanging, and 2 gauge boson according to the $t$-channel.
If $Z'$ exchanging works to explain CDF anomaly, $g'/m_{Z'}$ is 
huge like $g'/m_{Z'} \sim 500 $GeV.
Our CDM can interact with nuclei through $Z'$ exchanging, so that 
such large $g'/m_{Z'}$ leads so large direct cross section, 
$\sigma_{SI} \sim 0.01$pb. 
The mass of the CDM is constrained by the search for extra leptons, 
that is, it must be heavier than $100$ GeV,
where direct search for dark matters shows negative results.
Even if we consider the scenario that $Z'$ interaction is negligible 
in the direct scattering, for example,
in the case that only $(g^u_{R})_{ut}$ are large like the 3HDM,
$Z$ boson exchanging will work at the higher order. Such heavy CDM scenario 
is not favored by direct search.


Next, we discuss the case with extra $SU(3)$ triplets, 
$(q_{LI},\,\! q_{RI})$.\footnote{One can also replace them with fields charged 
under $SU(3)_c \times SU(2)_L$ with $1/6$ $U(1)_Y$ charges in this argument.} 
The mass terms are give by $\Phi$ like the masses of $l_{Li}$ and $l_{Ri}$.
Those extra colored particles would be stable because of the $U(1)'$ symmetry 
and $U(1)$ global symmetry corresponding to phase rotation 
of $q_{Li}$ and $q_{Ri}$.
In order to allow the decay of the extra colored particles, we introduce 
mixing term between $q_{L1,2}$ and $D_{Ri}$ according to adding $X$,
\begin{equation}
V_m = \lambda_i X^{\dagger} \overline{D_{Ri}} q_{L1} +\lambda_i X 
\overline{D_{Ri}} q_{L2},
\end{equation}
where $X$ is SM gauge-singlet scalar with $U(1)'$ charge, $Q_L$. 
If $X$ does not get nonzero vev, we can expect $X$ to be stable 
because of the remnant global symmetry after $U(1)'$ breaking. This type 
of dark matter has been well investigated in Ref.~\cite{Wise, Alwall:2010jc}.

It might be possible to consider charge assignments 
which only require the one extra generation. If the mixing 
between extra quarks and SM quarks is forbidden by $U(1)'$ charge,
we may need extra $SU(2)_L$ doublet scalar which has Yukawa coupling 
with extra quarks and SM quarks.
If the extra Higgs does not get vev, the neutral component could be 
a CDM candidate~\cite{Wise,Cirelli:2005uq}. 
This case also predicts large cross section through $Z$ and $Z'$ bosons.    

\section{Summary}
\label{sec:summary}
Let us summarize our results. 
In this paper, we constructed a complete $U(1)'$ model for flavor dependent 
couplings to the right-handed up-type quarks in the SM, and discussed 
the top FB asymmetry, the CDF $Wjj$ excess and cold dark matters. 
We first described the flavor dependent and leptophobic chiral $U(1)'$ models, 
and introduced new Higgs fields and fermions for the renormalizable Yukawa 
couplings for the SM fermions and the anomaly cancellation, respectively. 
Then the couplings of the SM fermions to the new $Z'$ and new flavored Higgs
doublets (in terms of the (pseudo) neutral/charged Higgs fields) 
were derived, and were used for the top FB asymmetry and the CDF 
$Wjj$ excess as well as the same sign top pair production at the LHC.  
We found that the interference effects between the $Z'$ and the Higgs bosons 
generally improves the overall description of the $t\overline{t}$ production 
cross section at the Tevatron and the top FB asymmetry, 
their $M_{t\overline{t}}$ distributions, 
and reduce the production cross section for the same sign top pair at the LHC. 
Our model for the top FB asymmetry can be tested in the near future at 
the Tevatron and the LHC by measuring $t\rightarrow Z' u$, 
$h \rightarrow t \bar{u} + c.c.$, and the single top production without  
the $b$ quark in the final states ($p \bar{p} , p p \rightarrow t Z'$). 
Also the spin-spin correlations of  $t\bar{t}$ and the longitudinal polarization 
of (anti)top quark \cite{Jung:2009pi,Jung:2010yn} could be  useful for testing 
our models based on the light $Z^{'}, h $ and $a$.  
It can not be too much emphasized that the  (pseudo) scalar Higgs bosons 
in our models are not added by hand, but should be included in the chiral 
$U(1)'$ flavor model of Ref.~\cite{Ko-Top}.
It is simply inconsistent to do phenomenology without them.
And most interestingly, these new Higgs doublets charged under $U(1)'$
not only help the models viable as a solution for the top FB asymmetry, 
but also can accommodate the CDF $Wjj$ excess for $\sin 2 \beta \sim 0.7$
through $p\bar{p} \rightarrow h^\pm \rightarrow W^\pm Z^{'}$, which 
is a kind of bonus when we made a completion of light leptophobic $Z'$ 
with flavor dependent couplings to the RH up-type quarks.
By making the light $Z'$ model for the top FB asymmetry mathematically 
consistent in terms of anomaly cancellation and physically realistic 
in terms of renormalizable Yukawa couplings, we found that the models 
come with new ingredients that could also accommodate other phenomena 
such as the CDF $Wjj$ excess or the CDM of the universe.
There are still constraints we did not touch in this paper, 
such as $\rho$ parameter.  In our models, extra Higgs doublets are charged 
under $U(1)'$ and induce the mixing between $Z$ and $Z'$. For example, 
the tree-level contribution will be written down by the form linear 
to $\sin \beta$ in 2HDM, 
so that our parameter choice must be consistent with the small mixing. 
We are sure that small $\sin \beta$, $( \lesssim O(10^{-1}))$, 
can realize large $(t,\,\! u)$ elements and enough large $Wjj$ excess 
without exceeding the observed $\rho$ parameter \cite{futurework}. 

Finally, we would like to note that our conclusions about extended 
multi Higgs doublets 
(with some of them being $U(1)^{'}$ flavored) and 
the related phenomenology would be generically true 
for many flavor gauge models with chiral couplings to the SM fermions.  
Some features obtained in this paper may be specific to our explicit models,  
depending on the new matter contents we introduce in order that we achieve 
the gauge anomaly cancellation and also allow the necessary Yukawa couplings. 
Our strategies can be adopted in any other 
attempts to construct realistic flavor models for the top FB asymmetry 
and the CDF $Wjj$ excess as well as $B_s -\overline{B_s}$ mixing in
terms of flavor changing $Z'$. 
This statement will apply to axigluon models, extra $W'$ model,
or flavor $SU(3)$ model for the right-handed up quarks 
(namely the models with chiral gauge interactions). 
It would not be enough to include only spin-1 vector bosons in order to 
discuss the top FB asymmetry or the same sign top pair productions. 
It is mandatory to construct the entire lagrangian including the 
realistic renormalizable Yukawa couplings, including new Higgs doublets 
that are charged under the chiral flavor gauge groups under consideration, 
in order to discuss the top physics.  New Higgs doublets 
that are charged under 
chiral gauge symmetry group can modify the top FB asymmetry, 
the same sign top pair productions and the $Wjj$. 
It will remain to be seen if the predictions on the top FB asymmetry, 
the same sign top pair and related phenomenology depends strongly on the 
additional Higgs doublets in addition to the original spin-1 gauge bosons
(axigluon, $Z'$, $W'$ or $SU(3)$ flavor gauge bosons) effects. 

\acknowledgments
We are grateful to Suyong Choi, Sunghoon Jung, Hyunsoo Kim, 
Soobong Kim, Sungwon Lee, Chang Seong Moon
Youngdo Oh and Un-ki Yang for useful discussions and communications.  
We thank Korea Institute for Advanced Study for providing computing resources 
(KIAS Center for Advanced Computation Abacus System) for this work.
The work of PK is supported in part by SRC program of National Research 
Foundation, Seoul National University, KNRC.
The work of CY is supported by Basic Science Research Program 
through the National Research Foundation of Korea(NRF) funded by the 
Ministry of Education Science and Technology(2011-0022996). 


\appendix
\section{Explicit description of 3 Higgs doublets model 
with $\bm{(-1,\,\! 0,\,\! 1)}$}
We show the explicit descriptions of gauge couplings and Yukawa couplings 
in the 3HDM of Sec. \ref{sec:3-Higgs}. As we see 
in the section, the couplings depend on the mixing, $(R_u)_{ij}$, 
and we have to avoid the FCNC constraint from $D^0$-$\overline{D}^0$ mixing, 
so that let us fix $R_u$ matrix as follows,
\begin{equation}
\label{eq:explicit-UR}
R_u=\begin{pmatrix} 
               e^{i \delta_1} &  0& 0   \\
                0 &   e^{i \delta_2} &  0 \\
               0 &  0 &  e^{i \delta_3} \end{pmatrix}\begin{pmatrix} 
               \cos \theta &  0& -\sin \theta   \\
                0 &   1 &  0 \\
               \sin \theta &  0 &  \cos \theta \end{pmatrix}.
\end{equation}

\subsection{Gauge couplings}
In this case, the matrix of gauge coupling, $g^u_R$, is 
\begin{equation}
  \label{eq:gR-3HDM}
g^u_R =\begin{pmatrix} 
           - \cos 2 \theta &  0& - e^{i(\delta_1 -\delta_3)} \sin 2\theta   \\
                0 &   0 &  0 \\
           - e^{-i(\delta_1 -\delta_3)}\sin 2\theta &  0 &   \cos 2 \theta 
       \end{pmatrix},
\end{equation}
and the others satisfy $g^d_{L,R}=g^u_L=0$. For this choice, there is no
$D^0 - \overline{D^0}$ mixing from $Z'$ exchanges.

\subsection{Yukawa couplings for neutral scalar Higgs}
Yukawa couplings for SM fermions and lightest neutral scalar Higgs, which are defined 
in the Eq. (\ref{eq:Yukawa-neutral}), are also described 
according to Eq. (\ref{eq:explicit-UR}),
\begin{eqnarray}
\left( Y^u_{ij} \right) &=&\begin{pmatrix} 
     \frac{m_u}{v} \left ( \frac{O_{11}^h \cos^2 \theta}{\cos \beta_1 \cos \beta_2} 
    +\frac{O_{31}^h \sin^2 \theta}{\sin \beta_2} \right) &  0& 
   \frac{m_ue^{i(\delta_1 -\delta_3)} }{2v}   
    \left ( \frac{O_{11}^h \sin 2 \theta }{\cos \beta_1 \cos \beta_2} 
    -\frac{O_{31}^h \sin 2 \theta }{\sin \beta_2} \right)  \\
          0 &   \frac{m_c}{v} \frac{O_{21}^h}{\sin \beta_1 \cos \beta_2} &  0 \\
                \frac{m_t e^{-i(\delta_1 -\delta_3)} }{2v}  
   \left ( \frac{O_{11}^h \sin 2 \theta }{\cos \beta_1 \cos \beta_2} 
   -\frac{O_{31}^h \sin 2 \theta}{\sin \beta_2} \right)  &  0 &  
   \frac{m_t}{v} \left ( \frac{O_{11}^h \sin^2 \theta}{\cos \beta_1 \cos \beta_2} 
   +\frac{O_{31}^h \cos^2 \theta}{\sin \beta_2} \right)   
     \end{pmatrix},  
\label{eq:Yuijm}
\nonumber \\
Y^d_{ij}  &=& \frac{m^d_i O_{21}^h}{v \sin \beta_1 \cos \beta_2 } \delta_{ij},
\nonumber \\
Y^e_{ij}  &=& \frac{m^l_i O_{21}^h}{v \sin \beta_1 \cos \beta_2 } \delta_{ij},
\end{eqnarray}
where $\beta_{1,2}$ are defined as 
\begin{equation}
( \langle H_1 \rangle,\,\! \langle H_2 \rangle,\,\! \langle H_3 \rangle)= 
\frac{v}{\sqrt{2}}(\cos \beta_1 \cos \beta_2,\,\! \sin \beta_1 \cos \beta_2,
\,\! \sin \beta_2  ).
\end{equation}
For this choice, there is no $D^0 - \overline{D^0}$ mixing from 
neutral Higgs exchanges.

\subsection{Yukawa coupling for charged Higgs and pseudo-scalar Higgs }
According to the orthogonal directions of Goldstone bosons, 
we could know the directions of the massive charged and pseudo-scalar Higgs. 
The direction of Goldstone mode will be 
$(\cos \beta_1 \cos \beta_2,\,\! \sin \beta_1 \cos \beta_2,\,\! \sin \beta_2)$, 
so that the other massive modes, $\xi_I (q=1,\,\! 2)$, will be written as
\begin{eqnarray}
\xi_1 &:&  (\cos \beta_1 \sin \beta_2,\,\! \sin \beta_1 \sin \beta_2,
\,\! -\cos \beta_2  ), \nonumber \\
\xi_2 &:&  (\sin \beta_1,\,\! -\cos \beta_1,\,\! 0  ).
\end{eqnarray}
The Yukawa couplings for the each charged Higgs and each pseudo-scalar Higgs 
in the Eq. (\ref{eq:Yukawa-charged}) and (\ref{eq:Yukawa-pseudo}) are
\begin{eqnarray}
\left( \frac{(V_{\textrm{CKM}})_{il} Y^{u1-}_{lj}}{\sqrt{2}} \right)
& =&\begin{pmatrix} 
        \frac{m_u \tan \beta_2}{v} \left (  \cos^2 \theta
       -\frac{ \sin^2 \theta}{\tan^2 \beta_2} \right) &  0& 
       \frac{m_u  \sin 2 \theta e^{i(\delta_1 -\delta_3)} }{v \sin 2 \beta_2 } 
\label{eq:chargedm}
  \\
        0 &   \frac{m_c \tan \beta_2}{v} &  0 \\
     \frac{m_t  \sin 2 \theta e^{-i(\delta_1 -\delta_3)} }{v \sin 2 \beta_2}
      &  0 &  \frac{m_t \tan \beta_2}{v} \left (  \sin^2 \theta 
       -\frac{\cos^2 \theta}{\tan^2 \beta_2} \right)   \end{pmatrix},   
\nonumber \\
Y^{d1+}_{ij} &=& \frac{ \sqrt{2} m^d_j \tan \beta_2}{v } \
(V_{\textrm{CKM}})_{ij},   
\end{eqnarray}
and for $\xi_2$ are
\begin{eqnarray}
\left( \frac{(V_{\textrm{CKM}})_{il} Y^{u2-}_{lj}}{\sqrt{2}} \right)
& =&\begin{pmatrix} 
      \frac{m_u \tan \beta_1}{v \cos \beta_2}   \cos^2 \theta &  0& 
      \frac{m_u \tan \beta_1 \sin 2 \theta}{2v \cos \beta_2 }
       e^{i(\delta_1 -\delta_3)}   \\
      0 &  - \frac{m_c }{v  \tan \beta_1 \cos \beta_2} &  0 \\
      \frac{m_t  \tan \beta_1 \sin 2 \theta }{2v \cos \beta_2}   
       e^{-i(\delta_1 -\delta_3)}  &  0 &  
     \frac{m_t \tan \beta_1}{v \cos \beta_2} \sin^2 \theta   
    \end{pmatrix},   \nonumber \\
Y^{d2+}_{ij}  &=& -\frac{\sqrt{2} m^d_j}{v  \tan \beta_1  \cos \beta_2  } 
(V_{\textrm{CKM}})_{ij}.             
\end{eqnarray} 
$Y^{au}_{ij}$ and $Y^{ad}_{ij}$ for each $\xi_q$ are given 
by $(V_{\textrm{CKM}})_{il}Y^{uq-}_{lj}/\sqrt{2}$ and $(V_{\textrm{CKM}})^{\dagger}_{il}Y^{dq+}_{lj}/\sqrt{2}$,
and $Y^{ae}_{ij}$ is given by replacing $m^d_j$ in the $(V_{\textrm{CKM}})^{\dagger}_{il}Y^{dq+}_{lj}/\sqrt{2}$ with $m^l_j$.

\subsection{$\bm{W^+_{\mu} \xi^-_q Z^{\prime \mu}}$ couplings}
We can describe the coupling of the charged Higgs with $W$ and $Z'$ bosons.
When we set the $U(1)'$ charges of Higgs, $(H_1,\,\! H_2,\,\! H_3)$, 
to $(q_{H1},\,\! q_{H2},\,\! q_{H3})$, the couplings of $\xi^{\pm}_q$ are
\begin{equation}
-V_{W^+_{\mu} \xi^-_1 Z'^{\mu}} = 2 m_W g' (q_{H1} \cos^2 \beta_1+q_{H2} 
\sin^2 \beta_1 -q_{H3} ) \sin \beta_2 \cos \beta_2 
(\xi_1^+W^-_{\mu} Z'^{\mu}+\xi_1^-W^+_{\mu} Z'^{\mu}),
\end{equation}
and
\begin{equation}
-V_{W^+_{\mu} \xi^-_2 Z'^{\mu}} = 2 m_W g' (q_{H1}-q_{H2}) 
\cos \beta_1 \sin \beta_1 \cos \beta_2 
(\xi_2^+W^-_{\mu} Z'^{\mu}+\xi_2^-W^+_{\mu} Z'^{\mu}).
\end{equation}

\end{document}